\documentclass[aps,showpacs,letterpaper,reprint,longbibliography,twocolumn,12point]{revtex4-1}

\usepackage{graphicx}% Include figure files
\usepackage{dcolumn}% Align table columns on decimal point
\usepackage{bm}% bold math
\usepackage{color}      % for coloring edit command

\usepackage{amstext}
\usepackage{amsmath}
\usepackage{graphics}
\usepackage{amssymb}
\usepackage{color}

\newcommand{\bra}[1]{\langle #1|}
\newcommand{\ket}[1]{|#1\rangle}

\newcommand \mcE{{\mathcal E}}

\newcommand \be{\begin{equation}}
\newcommand \ee{\end{equation}}
\newcommand \bea{\begin{eqnarray}}
\newcommand \eea{\end{eqnarray}}
\newcommand \nn{\nonumber}
\newcommand \opH {{\hat {\mathcal H}}}

\begin{document}

\title{A Rydberg blockade CNOT gate and entanglement in a 2D array of neutral atom qubits} 

\author{K. M.  Maller, M. T. Lichtman, T. Xia, Y. Sun}
\author{M. J. Piotrowicz}
\altaffiliation{Present address: Department of Physics, University of Michigan, Ann Arbor, MI 48109, USA}
\author{A.W.Carr}
\author{L. Isenhower}
\altaffiliation{Present address: Department of Physics, Abilene Christian University, Abilene, Texas 79699, USA}
\author{M. Saffman}

\affiliation{
Department of Physics, University of Wisconsin-Madison, 1150 University Avenue, Madison, Wisconsin 53706
}

\begin{abstract}
We present experimental results on two-qubit Rydberg blockade quantum gates and entanglement in a two-dimensional qubit array.   Without post selection against atom loss we achieve a Bell state fidelity of $0.73\pm 0.05$, the highest value reported to date. The experiments are performed in an array of single Cs atom qubits with a site to site spacing of $3.8~\mu\rm m$. Using the standard protocol for a Rydberg blockade C$_Z$ gate together with single qubit operations we create Bell states and measure their fidelity using parity oscillations. We analyze the role of AC Stark shifts that occur when using  two-photon Rydberg excitation and show how to tune experimental conditions for optimal gate fidelity. 
\end{abstract}

\date{\today}


\maketitle

\section{\label{sec:intro}Introduction}

Qubits encoded in hyperfine states of neutral atoms are a 
promising approach for scalable  implementations of quantum information processing\cite{Ladd2010}. We are developing an atomic qubit array for quantum logic experiments. The array consists of  qubits encoded in Cs atom hyperfine states. Single qubit gate operations are performed using either microwave fields for global operations on the array, or focused light fields for control of individual qubits\cite{Xia2015}.  Two-qubit entangling gates are based on Rydberg blockade interactions\cite{Jaksch2000}. Qubit initialization is performed with optical pumping and qubit readout is based on imaging of resonance fluorescence\cite{Saffman2005a}. 

Provided sufficiently high gate fidelities can be achieved the neutral atom approach provides a scalable path towards large qubit numbers. The qubit density in our recent 2D implementations\cite{Piotrowicz2013,Xia2015} is approximately one qubit per 14 $\mu\rm m^2$ with a loading fraction of 60\%. This translates into an effective area per qubit of 24 $\mu\rm m^2$. The area needed for a large number of qubits, say $10^6$, would be 
a modest 0.24 $\rm cm^2$. 
Although there are numerous engineering challenges associated with scaling to such a large number of qubits there is no fundamental reason why this could not be achieved. 

At the present time the largest impediment to scaling is that the demonstrated gate fidelities have not reached the level where fault-tolerant coding and error correction architectures are viable\cite{Devitt2013}. Single qubit gate operations have reached better than 0.99 fidelity with single site control\cite{Xia2015} and it is reasonable to anticipate further improvement using composite pulse sequences\cite{Merrill2014,*Mount2015}. The fidelity achieved to date for entangling gates is less satisfactory. Three research groups have demonstrated entanglement of neutral atom qubits using Rydberg interactions. The results have been characterized in terms of Bell state or entanglement fidelities with reported values of 0.58\cite{Isenhower2010}, 0.71\cite{Zhang2010}, 0.75\cite{Wilk2010}, and 0.81\cite{Jau2015} allowing for post selection to correct for atom loss during the gate sequence. Reported  fidelity results without post selection, which is preferable for quantum computing applications, are  0.58\cite{Zhang2010} and  0.60\cite{Jau2015}. 

The experimental entanglement fidelity results reported to date lag far behind theoretical analyses which predict gate fidelities $>0.99$\cite{XZhang2012} in a room temperature apparatus  and $>0.9999$ for circular Rydberg states at cryogenic 
temperatures\cite{Xia2013}.  It has therefore been an open question as to whether or not the separation between experimental and theoretical results is due to purely technical errors, or derives from some unaccounted for aspect of the atomic physics.  We demonstrate here that previous analysis has not fully accounted for AC Stark shifts that occur in two-photon excitation of Rydberg states. We clarify the impact of the Stark shifts on the effective gate matrix, and show how to minimize sensitivity to imperfectly controlled experimental parameters. We then demonstrate improved two-qubit entanglement with  fidelity of $0.79\pm0.05$ allowing for  post selection and $0.73\pm 0.05$  without post selection. Although still below what is needed for scalability we anticipate that further improvement will be possible in the future. 

The rest of the paper is organized as follows. In Sec. \ref{sec.ACStark} we recall the Rydberg blockade $C_Z$ protocol and analyze the impact of AC Stark shifts on the gate. We then proceed to show how to compensate for the Stark shifts to obtain an ideal gate matrix. In Sections \ref{sec.experiment}A, B we describe the experimental setup and how to measure relevant  parameters using Ramsey interference and Rabi oscillation experiments. In Sections \ref{sec.experiment}C,D  we describe CNOT and Bell state experiments, followed by a concluding section \ref{sec:discussion} with an outlook on future developments.

\section{Rydberg controlled phase gate}
\label{sec.ACStark}

\begin{figure}[!t] 
  \includegraphics[width=.95\columnwidth]{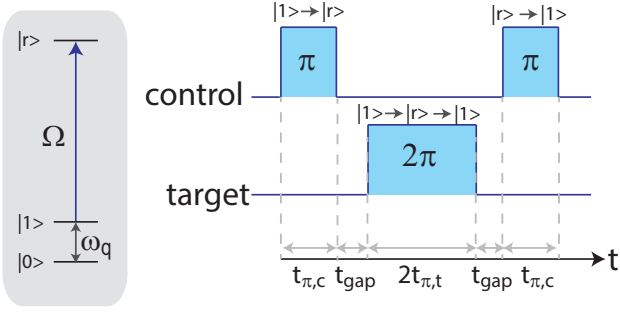}
  \caption{(color online) Level structure (left) and pulse sequence (right) for two-qubit 
Rydberg blockade $C_Z$ gate. The qubits are encoded in ground hyperfine states $\ket{0}, \ket{1}$ while $\ket{r}$ is a high lying Rydberg state. }
  \label{fig.setup}
\end{figure}

The standard protocol for creating a controlled phase gate via  Rydberg blockade uses a three pulse sequence to implement 
 \begin{equation}\label{eq:Czreg}
C_Z=\begin{pmatrix}
1&0&0&0\\
0&-1&0&0\\
0&0&-1&0\\
0&0&0&-1
\end{pmatrix}.
\end{equation}

The sequence  shown in Fig. \ref{fig.setup} uses a $\pi$ pulse on the control qubit, $2\pi$ on the target, and $\pi$ on the control\cite{Jaksch2000}.  In the ideal situation of perfect blockade and negligible ratio of excitation Rabi frequency $\Omega$ to qubit frequency splitting $\omega_{\rm q}$ we obtain the gate matrix (\ref{eq:Czreg}). While several  experiments have used single photon excitation of alkali atom Rydberg states\cite{Tong2004,*Manthey2014,*Hankin2014},\cite{Sassmannshausen2013}, all the Rydberg based quantum gate experiments except\cite{Jau2015} have used a more complicated two-photon excitation method. The primary reason for doing so has been to avoid the need for  high power at the short wavelengths of one-photon excitation (297 nm in Rb and 321 nm in Cs). 
As we proceed to show, the use of a two-photon drive changes the gate matrix so that even under ideal conditions we do not obtain Eq. (\ref{eq:Czreg}). 
Instead the $C_Z$  gate matrix takes the form 
\begin{equation}\label{eq:phasegate} 
C_{Z,\bar{\phi}}=\begin{pmatrix}
e^{\imath\phi_{00}}&0&0&0\\
0&e^{\imath\phi_{01}}&0&0\\
0&0&e^{\imath\phi_{10}}&0\\
0&0&0&e^{\imath\phi_{11}}
\end{pmatrix}
\end{equation} 
with $\bar\phi$ shorthand for the phases $\{\phi_{00},\phi_{01},\phi_{10},\phi_{11} \}.$
The $C_{Z,\bar\phi}$ operator can only create entanglement when $\phi_{00}-\phi_{01}-\phi_{10}+\phi_{11}\ne 2\pi n$ with $n$ integer. It is therefore essential to correctly control the gate phases. 

\renewcommand{\arraystretch}{1.5}
\begin{table*}[!ht]
\caption{Stark shifts  contributing to the gate phases. Here ${\mathcal E}_j$ is the field amplitude of field $j$, $\Omega_j$ is the Rabi frequency, $\Delta_1$ is the detuning from the intermediate level, and $\alpha_{g(R)j}$ is the non-resonant polarizability of the ground(Rydberg) levels at the frequency $\omega_j$ of field $j$.
Superscripts r and nr label resonant and nonresonant Stark shifts respectively.   Full expressions that account for the hyperfine structure of the intermediate level are given in Appendix \ref{sec.appendixA}}
\vspace{-.0cm}
\begin{center}
%\begin{ruledtabular}
\begin{tabular}{|l|c|c|c|}
\hline
description& shift on $\ket{0}$ & shift on $\ket{1}$& shift on $\ket{r}$\\
\hline
resonant shift from ${\mathcal E}_1$ &$\Delta_{01}^{\rm r}=\frac{1}{4}\frac{|\Omega_{1}|^2}{\Delta_1-\omega_{\rm q}}$   & $\Delta_{11}^{\rm r}=\frac{1}{4}\frac{|\Omega_{1}|^2}{\Delta_1}$&-\\
non-resonant shift from ${\mathcal E}_1$          & $\Delta_{\rm g1}^{\rm nr}=-\frac{1}{4\hbar}\alpha_{\mathrm{g}1}|{\mathcal E}_{1}|^2$ &  $\Delta_{\rm g1}^{\rm nr}$&$\Delta_{\rm R1}^{\rm nr}=-\frac{1}{4\hbar}\alpha_{\rm R1}|{\mathcal E}_{1}|^2$ \\
resonant shift from  ${\mathcal E}_2$     & -&  -  &$\Delta_{\rm R2}^{\rm r}=\frac{1}{4}\frac{|\Omega_{2}|^2}{\Delta_1}$\\
non-resonant shift from ${\mathcal E}_2$          & $\Delta_{\rm g2}^{\rm nr}=-\frac{1}{4\hbar}\alpha_{\mathrm g2}|{\mathcal E}_{2}|^2$ &  $\Delta_{\rm g2}^{\rm nr}$&$\Delta_{\rm R2}^{\rm nr}=-\frac{1}{4\hbar}\alpha_{\rm R2}|{\mathcal E}_{2}|^2$  \\
\hline
\end{tabular}
%\end{ruledtabular}
\end{center}
\label{tab:allshifts}
\vspace{-.8cm}
\end{table*}

\subsection{AC Stark shifts}

\begin{figure}[!t] 
  \includegraphics[width=.8\columnwidth]{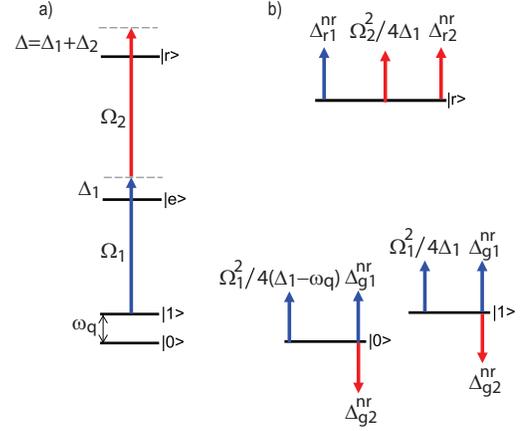}
  \caption{(color online) Level diagram (a) and ac Stark shifts (b) for two-photon Rydberg excitation. }
  \label{fig:rydlevel}
\end{figure}

The first contributions to the gate phases come from ac Stark shifts that arise due to the use of two-photon excitation as shown in Fig. \ref{fig:rydlevel}. We divide the Stark shifts into resonant, and  non-resonant contributions. Each of the qubit levels $\ket{0},\ket{1}$ acquire resonant and non-resonant Stark shifts as does the Rydberg level.  All shifts are  listed in Table \ref{tab:allshifts}. 

There are several things to note about the expressions given in  Table \ref{tab:allshifts}. The resonant shifts are the standard expressions valid for the situation where $|\Delta_1| \gg \gamma_e$ with $\gamma_e$ the radiative linewidth of the intermediate level. We assume two-photon resonance between  $\ket{1}$ and $\ket{r}$ so that 
$\Delta_2=-\Delta_1$ and $\Delta=\Delta_1+\Delta_2=0.$ The field amplitudes and Rabi frequencies are related by $\Omega_j = d_j {\mathcal E}_j/\hbar$ with $d_j$ the relevant transition dipole matrix element, $\mcE_j$ the electric field amplitude,  and the two-photon Rabi frequency  $\Omega_{\rm R}=\Omega_1\Omega_2/(2\Delta_1)$. In the approximation that the hyperfine splitting of  $\ket{e}$ is small compared to the detuning $\Delta_1$ the expressions given are valid. When this is not the case the expressions for the Stark shifts as well as the relation between $\Omega$ and the one-photon Rabi frequencies  have to be modified. The full expressions for  the specific case of Cs atoms excited via $6s_{1/2}\rightarrow 7p_{1/2}\rightarrow ns_{1/2}$, as in our experiments, are given in  Appendix \ref{sec.appendixA}. 

The non-resonant polarizabilities $\alpha_{gj}$ can be  calculated using a sum over states approach. Since we explicitly account for the resonant contributions to the ac Stark shift, the polarizabilities in this paper are defined with the resonant transitions excluded from the sum. The non-resonant Rydberg polarizability  is given by the expression 
\begin{eqnarray}
\alpha_{rj}=-\frac{e^2}{m_e\omega_j^2},
\label{eq.alpharyd}
\end{eqnarray}
where $e$ and $m_e$ are the electron charge and mass, respectively, and $\omega_j$ is the frequency of field $j$. We will assume that the excitation beams are large compared to the size of the Rydberg wavefunction and ignore corrections to the Rydberg shift arising from finite beam size effects\cite{SZhang2011}.

\subsection{Phase shift from a 2$\pi$ rotation}\label{sec.twopishift}

In the case of a two level system driven by a single field the ground state accumulates a shift of $e^{i\pi}=-1$ during  a resonant $2\pi$ pulse.  The situation is more complicated for the three level system driven by two fields.  The laser frequencies are tuned to give full population transfer from the ground state to the excited state.  This implies that the detuning compensates  the Stark shifts from each of the excitation beams.  In the case where there is two-photon resonance, or near resonance, and $\Delta_1/\gamma_{\rm e}$ is large to minimize spontaneous emission the resonance condition is
\begin{align}
\Delta\approx (\Delta_{\rm R2}^{\rm r}+\Delta_{\rm R1}^{\mathrm{nr}}+\Delta_{\rm R2}^{\mathrm{nr}})-\left(\Delta_{\rm 11}^{\rm r}+\Delta_{\rm g1}^{\mathrm{nr}}+\Delta_{\rm g2}^{\mathrm{nr}}\right)=0.
\end{align}
This includes the resonant Stark shift, which can be cancelled by setting $|\Omega_1|=|\Omega_2|$ (when the intermediate level hyperfine structure is negligible), and the  non-resonant shifts on the ground (Rydberg) states, $\Delta^{\mathrm{nr}}_{\rm g(R)1(2)}$.

In the far detuned limit when the hyperfine structure of the intermediate state can be neglected the phase accumulated by the ground state after a resonant $2\pi$ Rydberg pulse is
\begin{align}\label{eq:phyryd}
\phi_{\mathrm{R}}=\pi\left[1-\left|\frac{\Omega_{1}}{\Omega_{2}}\right|\mathrm{sign}(\Delta_1)\right]-(\Delta_{\rm g1}^{\mathrm{nr}}+\Delta_{\rm g2}^{\mathrm{nr}})2t_{\pi},
\end{align}
with $t_{2\pi}=2t_{\pi}=2\pi/|\Omega_{\rm R}|$. Equation \eqref{eq:phyryd} is readily derived using the Schr\"odinger equation for a three-level ladder configuration, and adiabatically eliminating the intermediate state in the limit of large detuning. 

We can express the Rabi frequency ratio as 
\bea
\pi\left|\frac{\Omega_{1}}{\Omega_{2}}\right|
&=&2\left|\frac{\pi2\Delta_1}{\Omega_1\Omega_2}\right|\frac{|\Omega_1|^2}{4\left|\Delta_1\right|}\nn\\
&=&{\rm sign}(\Delta_1)\frac{2\pi}{\left|\Omega_R\right|}\Delta_{11}^{\rm r}\nn\\
&=&{\rm sign}(\Delta_1)2t_\pi\Delta_{11}^{\rm r}\nn.
\eea
The $2\pi$ pulse  phase can therefore be written as 
\begin{align}\label{eq:phyryd2}
\phi_{\mathrm{R}}=\pi- (\Delta_{11}^{\rm r}+\Delta_{\rm g1}^{\mathrm{nr}}+\Delta_{\rm g2}^{\mathrm{nr}})2t_{\pi}.
\end{align}

This way of writing the phase has a clear physical interpretation. The factor of $\pi$ is the quantum phase accumulation from rotating the effective two-level system, which is analogous to a spin $1/2$, through $2\pi$. The second term is the Stark phase accumulated by the ground state over a time of  $2t_\pi$. 

It may be surprising that the usual picture of a $2\pi$ Rabi pulse imparting a $\pi$ phase shift is only valid for a single photon transition. For a two-photon 
drive the phase shift can take on any possible value although values of the Rabi frequencies for which $\phi_{\mathrm{R}}=\pi$ can always be found.
In particular   when the ac Stark shifts on the ground state are fully compensated,
i.e. $\Delta_{11}^{\rm r}+\Delta_{\rm g1}^{\rm nr} + \Delta_{\rm g2}^{\rm nr}=0$, then $\phi_{\rm R}=\pi$, and we recover the 
one-photon transition result.

\subsection{Gate phases}
\label{sec.phasecalcs}

The 
$C_{Z,\bar\phi}$ operator is constructed by considering how the shifts described in the previous sections affect the computational basis states, $|c t\rangle=\{|00\rangle,|01\rangle,|10\rangle,|11\rangle\}$.  

For the $|00\rangle$ state the excitation beams are detuned by $\omega_{\rm q}$ for both qubits so both remain in the $|00\rangle$ state.  From Table \ref{tab:allshifts}, the shifts on $|0\rangle$ include the resonant ${\mathcal E}_1$ shift and the non-resonant ${\mathcal E}_1$ and ${\mathcal E}_2$ shifts.  The shift on $|0\rangle$ for the control(target) qubits is 
\begin{align}
\phi_{\mathrm{hf},\rm c(t)}=-\left(\Delta_{01,\rm c(t)}^{\rm r}
+\Delta^{\mathrm{nr}}_{\rm g1,c(t)}+\Delta^{\mathrm{nr}}_{\rm g2,c(t)}\right)2
t_{\pi,\rm c(t)},
\end{align}
so
\begin{align}
\phi_{00}=\phi_{\mathrm{hf},\rm c}+\phi_{\mathrm{hf},\rm t}.
\label{eq.phi00}
\end{align}
Here we have allowed for a possible variation in parameters at the control and target qubit sites so that $\phi_{\mathrm{hf},c}$  $\phi_{\mathrm{hf},t}$  need not be equal . 

For the $|01\rangle$ state, the control experiences the off-resonant Stark shift, $\phi_{\mathrm{hf},c}$, while the target picks up a phase shift from the resonant $2\pi$ Rydberg pulse, $\phi_{\mathrm{R}}$ from Eq. \eqref{eq:phyryd},
\begin{align}
\phi_{01}=\phi_{\mathrm{hf},\rm c}+\phi_{\mathrm{R},\rm t}.
\label{eq.phi01}
\end{align}

The $|10\rangle$ state is different than the $|01\rangle$ state because the control atom is held in the Rydberg state for a time $2t_{\rm gap}+2t_{\pi,\rm t}$.  The phase accumulated during this time is   due to  the ground-Rydberg differential Stark shift and is given by 
\begin{eqnarray}
\phi_{\mathrm{gap}}&=&-\left[\left(\Delta_{\rm R2,\rm c}^{\rm r}+\Delta^{\mathrm{nr}}_{\rm R1,c}+\Delta^{\mathrm{nr}}_{\rm R2,c}\right)\right.\nonumber\\
&-&\left. \left(\Delta_{\rm 11,\rm c}^{\rm r}+\Delta^{\mathrm{nr}}_{\rm g1,c}+\Delta^{\mathrm{nr}}_{\rm g2,c}\right)\right]2\left(t_{\mathrm{gap}}+t_{\pi,\rm t}\right).
\end{eqnarray}
Here $t_{\mathrm{gap}}$ is the extra time in between pulses which, experimentally, is the minimum time it takes to switch the laser beams between control and target sites.  In total the shift on the $|10\rangle$ state is 
\begin{align}
\phi_{10}=\phi_{\mathrm{R},\rm c}+\phi_{\mathrm{gap}}+\phi_{\mathrm{hf}, \rm t}.
\label{eq.phi10}
\end{align}

Finally the $|11\rangle$ state experiences an additional shift due to the blockade, $\phi_{\mathrm{B}}$, for a time $2t_{\pi, \rm t}$ which includes the resonant Stark shift from ${\mathcal E}_1$ and the non-resonant shifts on the $|1\rangle$ state of the target atom
\begin{eqnarray}
\phi_{\mathrm{B}}&=&-\left(\Delta_{11,\rm t}^{\rm r}+\Delta^{\mathrm{nr}}_{\rm g1,t}+\Delta^{\mathrm{nr}}_{\rm g2,t}\right)2t_{\pi,\rm t}+\phi_{\rm BL},\nn\\
&=&-\pi +  \phi_{\rm R,t}+\phi_{\rm BL}.\nn
\end{eqnarray}
The last term is a small blockade leakage phase\cite{Jaksch2000} $\phi_{\rm BL}=\pi\Omega_{\rm R}/(2 {\sf B}).$
The total  phase accumulation on the $|11\rangle$ state during the gate sequence is thus 
\begin{align}
\phi_{11}=-\pi+ \phi_{\mathrm{R},\rm c}+\phi_{\mathrm{R},\rm t}+\phi_{\mathrm{gap}}+\phi_{\mathrm{BL}}.
\label{eq.phi11}
\end{align}
 Equations (\ref{eq.phi00}-\ref{eq.phi11}) fully determine the phases of the $C_{Z,\bar\phi}$ operator which we summarize here for convenience
\begin{eqnarray}
\phi_{00}&=&\phi_{\mathrm{hf},\rm c}+\phi_{\mathrm{hf},\rm t},\nn\\
\phi_{01}&=&\phi_{\mathrm{hf},\rm c}+\phi_{\mathrm{R},\rm t},\nn\\
\phi_{10}&=&\phi_{\mathrm{R},\rm c}+\phi_{\mathrm{gap}}+\phi_{\mathrm{hf}, \rm t},\nn\\
\phi_{11}&=&-\pi+ \phi_{\mathrm{R},\rm c}+\phi_{\mathrm{R},\rm t}+\phi_{\mathrm{gap}}+\phi_{\mathrm{BL}}.\nn
\end{eqnarray}

The gate phases are not completely independent since 
$$
\phi_{01}+\phi_{10}-\phi_{00}-\phi_{11}=\pi-\phi_{\mathrm{BL}}.
$$
In the limit where the blockade leakage phase $\phi_{\mathrm{BL}}$ is small, which will be the case for parameters which yield high fidelity entanglement,  there is a fixed constraint between the phases. A global multiplicative phase factor is irrelevant, leaving two free phases. As we discuss in the following section a correct choice of two  parameters is sufficient to fix the two phases and obtain an ideal gate operation. 

\subsection{Setting parameters to recover an ideal $C_Z$ gate}\label{sec:parmcomp}

In general the $C_{Z,\bar\phi}$ operator does not necessarily create entanglement, and does not directly create Bell states with standard phases. In this section we show that it is in principle possible to choose parameters such that an ideal $C_Z$ operator is implemented by the pulse sequence of Fig. \ref{fig.setup}.  For simplicity we assume that $\Omega_1, \Omega_2$ are the same for both control and target atoms so that we can drop the $\rm c,t$ subscripts on the gate phases.  We will also neglect $\phi_{\rm BL}$ since it is a small error for typical experimental parameters of ${\sf B}\gg \Omega_{\rm R}$. Alternatively the $\phi_{\rm BL}$ phase can be cancelled using a slightly modified pulse sequence, which does not change the other gate phases, as described in Fig. 3 of Ref. \cite{XZhang2012}.

To proceed we note that we can always ensure $\phi_{\mathrm{R}}=n\pi$ by choosing the correct value for the Rabi frequency ratio $q=\left|\frac{\Omega_2}{\Omega_1}\right|$. To see this let 
$\Delta_{11}^{\rm r}=a |\Omega_1|^2,$ 
$\Delta_{g1}^{\rm nr}=b |\Omega_1|^2,$ 
$\Delta_{g2}^{\rm nr}=c |\Omega_2|^2,$ 
$t_\pi=\pi d/|\Omega_1 \Omega_2|$ with $a,b,c,d$ real constants that depend on the detuning from the intermediate level and atomic structure parameters. If we then choose $q$ such that  
$$
1-n= \frac{(a+b)d}{q}+(c d) q
$$
we obtain $\phi_{\mathrm{R}}=n\pi$.
We then set  $t_{\rm gap}$  such that $\phi_{\mathrm{gap}}=2n'\pi$.  Solutions occur at 
\begin{align}
\begin{split} &t_{\mathrm{gap}}+t_{\pi}=\\&\frac{n'\pi}
{\left(\Delta_{\rm R2}^{\rm r}+\Delta^{\mathrm{nr}}_{\rm R1}+\Delta^{\mathrm{nr}}_{\rm R2}\right)-
\left(\Delta_{11}^{\rm r}+\Delta^{\mathrm{nr}}_{\rm g1}+\Delta^{\mathrm{nr}}_{\rm g2}\right)}\end{split}.\nn
\end{align}

Setting $\phi_{\rm R}, \phi_{\rm gap}$ to be multiples of $\pi$ as described above, the gate phases modulo $2\pi$, 
for $n$ odd and $n'$ even are 
\begin{align}
\begin{split} 
\phi_{00}=&2\phi_{\mathrm{hf}},\\
\phi_{01}=&\phi_{\mathrm{hf}}+\pi,\\
\phi_{10}=&\phi_{\mathrm{hf}}+\pi,\\
\phi_{11}=&\pi.\end{split}
\end{align}  
For $n$ even and $n'$ odd we get 
\begin{align}
\begin{split} 
\phi_{00}=&2\phi_{\mathrm{hf}},\\
\phi_{01}=&\phi_{\mathrm{hf}},\\
\phi_{10}=&\phi_{\mathrm{hf}},\\
\phi_{11}=&\pi.\end{split}
\end{align} 
The $\phi_{\mathrm{hf}}$ phases can be corrected by applying global $R_z(\theta)$ rotations with $\theta=-\phi_{\mathrm{hf}}$ which recovers the  ideal $C_Z$ of Eq.  \eqref{eq:Czreg}.  We emphasize that an ideal $C_Z$ gate is recovered apart from errors due to spontaneous emission from the intermediate and Rydberg levels, finite temperature Doppler and position fluctuations, and finite blockade strength. Those errors have been quantitatively studied in previous work\cite{XZhang2012}, but without the constraints implied by the parameter choices presented here. We defer a reexamination of the theoretically achievable  gate fidelity in a real atom to future work.

\subsection{Setting parameters to recover a $C_X$ gate}
\label{sec.CXgate}

In the experiments described below in Sec. \ref{sec.experiment} we have not implemented the parameter settings needed for an ideal $C_Z$ gate. Nevertheless we can still create a modified $CNOT=C_X$ gate and create Bell states, albeit not with the standard phases. The standard $C_X$ gate in the computational basis $\{00,01,10,11\}$ is
$$
\mathrm{C}_{X}=\begin{pmatrix}
1&0&0&0\\
0&1&0&0\\
0&0&0&1\\
0&0&1&0
\end{pmatrix}.$$
An equivalent operator, but with the $X$ operation conditioned on the control qubit being in state $|0\rangle$, is  
$$
\mathrm{\bar C}_{X}=(X\otimes I)\mathrm{C}_{X}(X\otimes I)=\begin{pmatrix}
0&1&0&0\\
1&0&0&0\\
0&0&1&0\\
0&0&0&1
\end{pmatrix}.$$

The usual method to transform the $C_Z$ gate into a $C_X$ gate is to apply Hadamard gates on the target qubit before and after the $C_Z$ operation.  We generalize this to $\pi/2$ rotations, with a relative phase $\theta$ to create the operator 
$$
C_{X,\bar\phi}(\theta)=R_{\rm t}(\pi/2,\theta)C_{Z,\bar\phi}R_{\rm t}(\pi/2,0)
$$ 
 where $R_{\rm c(t)}(\xi,\theta)$ is a ground state rotation on the control(target) by  an angle $\xi$, with phase $\theta$. For arbitrary $\bar\phi$ and $\theta$ we find 
$$
\mathrm{C}_{X,\bar\phi}(\theta)=\begin{pmatrix}
a&b&0&0\\
c&d&0&0\\
0&0&e&f\\
0&0&g&h
\end{pmatrix}$$
with eight nonzero elements $a-h$. However, for specific values of $\theta$ we get a $C_{X,\bar\phi}(\theta)$ operator proportional  to $\mathrm{C}_{X}$
or $\mathrm{\bar C}_{X}$, with only four nonzero elements of unit modulus, but with different  phase factors. We define a matrix overlap  as 
 $O(\mathrm{C}_{X},\mathrm{C}_{X, \bar\phi}(\theta))\equiv\frac{1}{4}\sum\limits_{\rm elements}\mathrm{C}_{X}|\mathrm{C}_{X,\bar\phi}(\theta)|^2$.  
It then follows that 
\begin{align}\label{eq:realcnots}
\begin{split}
O(\mathrm{C}_{X},\mathrm{C}_{X,\bar\phi}(\phi_{11}-\phi_{10}))=1,\\
O(\mathrm{\bar C}_{X},\mathrm{C}_{X,\bar\phi}(\phi_{01}-\phi_{00}))=1.
\end{split}
\end{align}
Note that the overlap is not a gate fidelity and does not contain any phase information.  
The difference between the phases used to prepare the two $C_X$ gates is independent of  parameters as expected,
\begin{align}\begin{split}
(\phi_{11}-\phi_{10})-(\phi_{01}-\phi_{00})
&=-\phi_{\mathrm{R},t}+\phi_{\mathrm{B}}\\
&=-\pi.
\end{split}\end{align}

We then  use the $C_{X,\bar\phi}(\theta)$ gate to create Bell like states, but with nonstandard phases. To do so we apply the sequence
 \begin{eqnarray}\label{eq:bellstark}
U(\theta,\bar\phi)&=&C_{X,\bar\phi}(\theta)R_c(\pi/2,0).
\end{eqnarray}
Operating with $U$ on the product state $\ket{00}$ creates a maximally entangled state when  $\theta=\phi_{11}-\phi_{10}$. In general this will not be one of the Bell states due to the presence of different phase factors on the two components of the state vector. If desired we can recover a standard Bell state with additional one qubit rotations.  This is the approach we demonstrate in the next section.

To determine the entanglement fidelity of the state created with $U$  we  measure the populations and the two-qubit coherence terms which can  be done by the method of  parity oscillations\cite{Turchette1998}.  Defining the parity signal $P=P_{00}+P_{11}-P_{01}-P_{10}$, where $P_{ij}$ are the diagonal terms of the two qubit density matrix,  a $\pi/2$ rotation at an angle $\theta$ on both qubits transforms  the parity signal to
 \begin{align}
P'=2{\rm Re}[C_2]-2|C_1|\cos(2\theta+\phi).
\end{align}
Here $C_1=|C_1|e^{\imath\phi}$ is the coherence between states $\ket{00}$ and $\ket{11}$ and $C_2$ is the coherence between states $\ket{01}$ and $\ket{10}$. 
The coherence $C_1$ can then be extracted from the parity oscillation measurements.  
The entanglement fidelity of states close to the Bell state $\frac{\ket{00}+\ket{11}}{\sqrt2}$ is   $F=\frac{P_{00}+P_{11}}{2}+|C_1|$. Values of $F>0.5$ are a sufficient condition for the presence of entanglement\cite{Sackett2000}.

\section{Experiment}
\label{sec.experiment}

\subsection{Set-Up}

Cs atoms are loaded into a 49 site array of blue-detuned dipole traps formed by 64 tightly focused, weakly overlapping 780~nm beams as described in Refs.  \cite{Piotrowicz2013,Xia2015}.  The $7\times7$ site array has a $3.8~\mu$m site to site spacing.  The qubits are encoded in the  hyperfine clock states with $|0\rangle\equiv|6s_{1/2}, f = 3, m_f = 0\rangle$ and $|1\rangle\equiv|6s_{1/2}, f = 4, m_f = 0\rangle$.  Single qubit rotations are performed globally using a 9.2~GHz microwave field. Single site rotations use a combination of the microwave field and a tightly focused beam detuned by -14~GHz from the $7p_{1/2}$ line to induce a differential Stark shift equal to $\sim40$~kHz on a single site. In contrast to the approach of Ref. \cite{Xia2015} where the microwave frequency was detuned from $\omega_{\rm q}$ and the optical Stark shift provided a local resonance condition, here the microwave frequency is set to $\omega_{\rm q}$ and the Stark shift is used to tune the site where no rotation is desired out of resonance. 

Rydberg excitations are driven using a two-photon transition through an intermediate $7p_{1/2}$ level.  The two wavelengths  are 459~nm and 1038~nm.  The lasers are locked to high finesse ULE resonators for long term frequency stability and short term linewidths $<500$ Hz on few $\mu\rm s$ timescales.  We measure $\Delta_1$, the detuning from the center of mass of $7p_{1/2}$,  by maximizing the light scattering from the $|7p_{1/2},f=4\rangle$ level, and then detuning the light by a known amount. The hyperfine constant of $7p_{1/2}$ is reported in Table \ref{tab.CsRyd1}.  The duration of the excitation pulses is controlled by acousto-optic modulators (AOMs). The pulses were of 
square shape as indicated in Fig.   \ref{fig.setup},
with typical rise and fall times of  50 ns. 
For each Rydberg pulse the 1038 nm light was left on for approximately 50 ns longer than the 459 nm light so that the precise value of the pulse duration was controlled by the 459 nm light.  This leads to some additional ground-Rydberg AC Stark shifts from the 1038 nm light that are not accounted for in the analysis of Sec. \ref{sec.phasecalcs}.

\begin{figure*}[!t]
\includegraphics[width=17.cm]{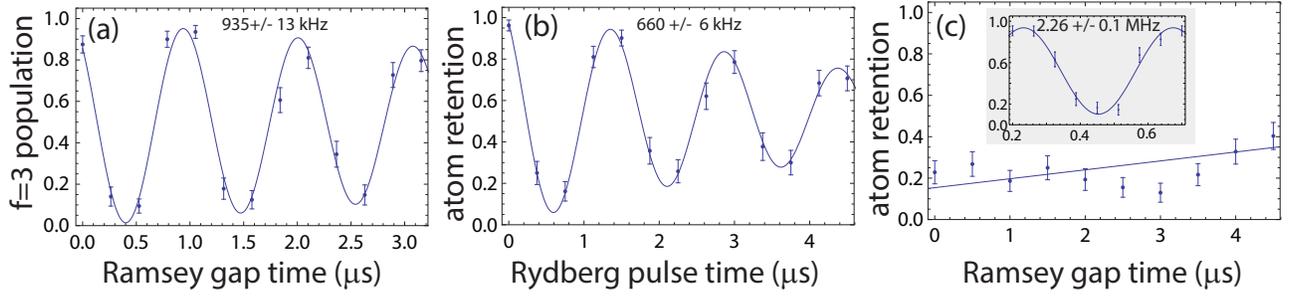}
\vspace{-.0cm}
\caption{(color online) Experimental data used to determine optical beam powers. The figure shows measurements on the control site only and fitted frequencies are indicated in the plots. (a) ground state Ramsey oscillations  with 459 nm laser used to extract $\Omega_1$. (b) ground-Rydberg Rabi oscillations used to extract $\Omega_2$ once $\Omega_1$ and $\Delta_1$ are known.   (c) ground-Rydberg Ramsey experiment used to measure $\Delta_{\rm diff,gR}$.  The straight line is an aid to the eye and is not a fit, but suggests a frequency $< 0.1~\rm MHz$.  The inset shows a faster ground-Rydberg Ramsey oscillation using other parameters. Beam powers  extracted from these measurements are given in Table \ref{tab.expparameters}.}%(a) Experimental data from $2014\_10\_27\_11\_08\_20$.(b)  Data from $2014\_11\_03\_17\_05\_32$. (c) Data from $2014\_11\_12\_16\_30\_07$.
  \label{fig:ryd82S22GHz}
\label{fig.array}
\end{figure*}
 Each beam is sent through separate fibers to a 2D beam scanner which is created using two crossed AOMs which allows for 2D positioning across the array \cite{Maller2015b}. The two counter-propagating  beams are focused to waists ($1/e^2$ intensity radii) of $3.0$ and $3.7~\mu\rm m$ for the 459 and 1038 nm light, and aligned onto a single site.  The  site to site switching time is $\sim0.5$~$\mu$s.
The two photons are $\sigma_+, \sigma_-$ polarized with respect to the quantization axis $z$ which is perpendicular to the plane of the qubit array. A 0.15 mT bias magnetic field is applied along $z$. With these polarizations and choice of intermediate level we excite a Rydberg $|ns_{1/2},m_j=-1/2\rangle$ fine structure state. As shown in Appendix \ref{sec.Rabi2fs} there is negligible excitation of the $m_j=+1/2$ Zeeman state. All data reported here are for the $|82s_{1/2},m_j=-1/2\rangle$ state with $\Delta_1=2\pi\times 0.83 ~\rm GHz$.  The optical trap array is turned off  for the few $\mu\rm s$ duration of the Rydberg gates. Turning the traps back on after Rydberg excitation leads to photoionization or mechanical loss of the atoms before they decay to the ground state. In this way trap loss is used to measure the Rydberg excitation probability.   

As we discuss in the next section the detuning $\Delta_1$ and the beam powers are chosen to minimize the differential ac Stark shift between ground and Rydberg states which can be expressed as 
\begin{eqnarray}
\Delta_{\mathrm{diff},\rm gR}&=&\left(\Delta_{\rm R2}^{\rm r}+\Delta^{\mathrm{nr}}_{\rm R1}+\Delta^{\mathrm{nr}}_{\rm R2}\right)- \left(\Delta_{\rm 11}^{\rm r}+\Delta^{\mathrm{nr}}_{\rm g1}+\Delta^{\mathrm{nr}}_{\rm g2}\right).\nn
\end{eqnarray}
 Minimizing this shift is advantageous as it reduces sensitivity to intensity fluctuations caused by laser instability, optical beam pointing drifts, and atomic motion. A small differential Stark shift also prevents time varying detuning, and consequently off-resonant state rotations,  during the finite rise and fall times of the optical pulses.    Because the experimental detuning is only several times larger then the $7p_{1/2}$ hyperfine splitting, the hyperfine structure must be included when calculating the differential shift (see Appendix \ref{sec.appendixA}).  State dependent Stark shifts due to the 780 nm trapping light\cite{SZhang2011} are not accounted for since we turn off the traps for the few $\mu\rm s$ duration of the Rydberg gates.

\subsection{Setting and Estimating parameters}\label{sec:parmestimate}

In order to calculate all of the shifts induced by the Rydberg excitation beams we need to have good measurements of the beam intensities at the atoms.  A series of Ramsey and Rabi flopping experiments is used for this purpose as shown in Fig. \ref{fig.array}. 

To find  $\mcE_1$, two microwave $\pi/2$-pulses are applied with a variable length 459~nm pulse in between.  The resulting Ramsey frequency is equal to the differential Stark shift of the qubit states induced by the 459~nm light. This is equal to the difference between the resonant Stark shifts on $|1\rangle$ and $|0\rangle$ which are separated by the hyperfine frequency $\omega_{\rm q}$,  
\begin{align}\label{eq:diff459}
\Delta_{\mathrm{diff},g}=\Delta_{11}^{\rm r}-\Delta_{01}^{\rm r}.
\end{align}

\begin{table*}[!ht]
\centering
\begin{tabular}{|l|c|c|c|c|c|}
\hline
measured values  & range over 2 days& value used for fits& value used for fits & $C_X$ eye diagram \\
  & (MHz)& control site (MHz)& target site (MHz)& \\
\hline
$\Delta_{\rm diff,g}/2\pi$  &0.80-0.95 &0.86   &0.81 &\\
$\Omega_{R}/2\pi$&.63-.75  & 0.67& 0.65 &\\
$\Delta_{\rm diff,gR}/2\pi$ &$\stackrel{<}{\sim}0.1$ & &&  \\
$\phi_{01}-\phi_{00}$ & &  & &   1.05 (rad)\\
\hline
inferred values  &  & &  &\\
\hline
$\Delta_{\rm inferred,gR}/2\pi$ &  & 0.088&0.160  & \\
$\phi_{01}-\phi_{00}$ & & &   & 1.01  (rad)  \\
\hline
\end{tabular}
\caption{Experimentally determined optical parameters. The first  column give the range of measured values from multiple measurements over a two day time span. The next two columns give the values assumed for determining the optical beam powers. Using beam waists of $w_{459}=3.0~\mu\rm m$, $w_{1038}=3.7~\mu\rm m$ the fitted powers were $P_{459,c}= 22.~\mu\rm W,$ $P_{459,t}= 21.~\mu\rm W,$ $P_{1038,c}= 1.9~\rm mW,$ and  $P_{1038,t}=2.0~\rm mW $. The inferred ground-Rydberg differential Stark shift and $C_X$ phase using these beam powers are given in the last two rows. }
\label{tab.expparameters}
\end{table*}

Measuring $\Delta_{\mathrm{diff},g}$, and using the dependence on the intensity of the 459 nm light we can infer $\mcE_{1}$ since  $\Delta_1$ and $\omega_{\rm q}$ are known.  In principal the same method can be used to extract $\mcE_{2}$ but the differential shift on the ground state from the 1038~nm beam is small, about 200~Hz.  Instead we measure $\Omega_{\rm R}$ the  two-photon Rabi frequency for Rydberg excitation. Since $\Omega_{\rm R}$ depends on $\mcE_1, \mcE_2,\Delta_1,$ and the hyperfine structure of the intermediate level, which is known, we can infer $\mcE_2$. 
We emphasize that for our experimental parameters it is important 
to account for the hyperfine structure using the expressions given in Appendix \ref{sec.appendixA}. Although the full expressions only differ by about 10\% from the approximate expressions of Table \ref{tab:allshifts} the gate performance is very sensitive to the beam intensities at the 10\% level, as can be seen in Fig. \ref{fig:EntangleSens} below.

 As a consistency check we then measure the ground-Rydberg differential shift $\Delta_{\rm diff,gR}$ with a ground-Rydberg Ramsey measurement.  This shift depends on both $\mcE_1$ and $\mcE_2$. Figure \ref{fig:ryd82S22GHz}   shows measured data  using the Rydberg 82$s_{1/2}$ state. The measured  and inferred quantities  are listed in 
Table \ref{tab.expparameters}. 
 Using the  matrix elements from Table \ref{tab.CsRyd1} we determine the field strengths $\mcE_j$ which are used to calculate resonant and non resonant Stark shifts which in turn are used to infer $\Delta_{\rm inferred,gR}$. Figure  \ref{fig:ryd82S22GHz}(c) shows the ground-Rydberg differential shift after choosing beam powers such that the shift is relatively small, less than 100 kHz. At these low frequencies we are not able to measure the shift accurately, since we cannot hold the Rydberg atoms for extended periods. We therefore only give an estimated upper limit on the shift in the Table.

Note the parameters above are not fine tuned to recover an ideal $C_Z$ as discussed in section \ref{sec:parmcomp}. We instead use a relative phase of  $\phi_{01}-\phi_{00}$ between the ground state $\pi/2$ rotations to recover an entangling $C_{X,\rm tc}$ gate, as explained in Sec. \ref{sec.CXgate}.
With the measured beam parameters, we  use the equations of Sec. \ref{sec.phasecalcs} to calculate the expected   two qubit operators
\begin{align}\label{eq:exampleCZ}
C_{Z,\bar\phi}=\left(
\begin{array}{cccc}
 e^{-\imath\,  0.13} & 0 & 0 & 0 \\
 0 & e^{\imath\,  0.88} & 0 & 0 \\
 0 & 0 & e^{-\imath\, 0.76} & 0 \\
 0 & 0 & 0 & -e^{\imath\,  0.24 } \\
\end{array}
\right),\end{align} and 
\begin{align}\label{eq:exampleCnot}
C_{X,\bar\phi}(\phi_{01}-\phi_{00})=\left(
\begin{array}{cccc}
 0 & e^{-\imath 1.7} & 0 & 0 \\
 e^{-\imath 0.69} & 0 & 0 & 0 \\
 0 & 0 & e^{-\imath0.76} & 0 \\
 0 & 0 & 0 &  e^{-\imath 2.89} \\
\end{array}\right).\end{align}
All nonlisted elements in $C_{X,\bar\phi}$ have magnitude $<8\times 10^{-4}$. 
We see that the gate phases differ appreciably from the standard values.
Nonetheless we can still create entangled states as explained in Sec. \ref{sec.CXgate}. The predicted Bell type state produced by this gate is 
\begin{eqnarray}
\ket{\psi}&=&-e^{\imath\phi_{00}} \frac{\ket{00}+e^{\imath(-2\phi_{00}+\phi_{01}+\phi_{10})}\ket{11}}{\sqrt2}\nonumber\\
&=&-e^{\imath\phi_{00}}\frac{\ket{00}+e^{\imath0.38}\ket{11}}{\sqrt2}\nonumber
\end{eqnarray}
 which is a maximally entangled state.  The largest effect not accounted for in the predicted state is the finite blockade strength of the two-qubit Rydberg interaction. We use a  Rydberg excitation Rabi frequency of  $\Omega_{\rm R}=2\pi\times .67~\rm MHz$. The Rydberg interaction for $82s_{1/2}$ states at $7.6~\mu\rm m$ separation is  ${\sf B}\simeq 2\pi\times 23~\rm MHz$. We calculate a blockade leakage phase of $\phi_{\rm BL}=\frac{\pi\Omega}{2 {\sf B}}\simeq$ 2.6 deg., which is negligible compared to other error sources.

\subsection{$C_X$ gate-experimental results}\label{sec.cnoteye}

\begin{figure}[!t] 
  \centering
  \includegraphics[width=.99\columnwidth]{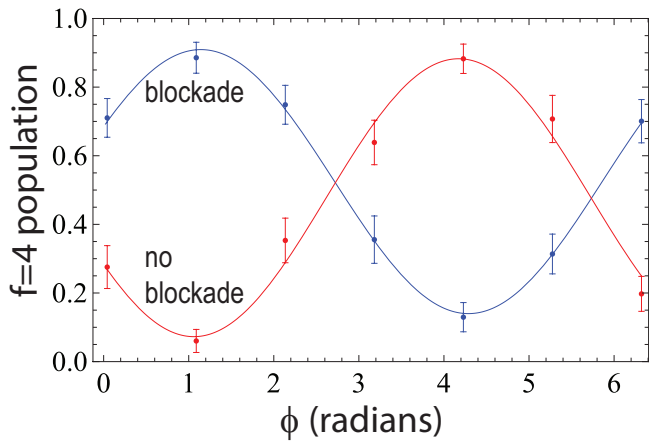}
   \caption{(color online) Experimental $C_X$ eye diagram on next nearest neighbor sites separated by $7.6~\mu\rm m$.  The blue curve is from loading an atom in both control and target sites and represents a blockaded data set.  The red curve is from data with no atom in the control site.  The two curves have a $\pi$ phase shift with respect to each other as expected.   The phase at the minimum of the no blockade curve is $\phi_{01}-\phi_{00}$ which gives the $C_{X,\rm tc}$ gate.  %Experimental data from $2015_04_24_14_35_43
}
  \label{fig:cnoteye}
\end{figure}

It is clear that setting the correct phase of the second ground state pulse is crucial to the operation of the gate and creation of entangled states.  This phase is found experimentally by varying the phase of the final ground state pulse. An example of this can be seen in Fig. \ref{fig:cnoteye} for control and target qubits that are two sites away so that their separation is 7.6~$\mu$m.  Single atom data of the target atom which is cut on whether a control atom is present or not is shown.  The blue curve shows the data when a control atom is present and therefore the Rydberg blockade occurs.  While this is only single atom data, this curve can be thought of as representing the $|11\rangle$ state and if we wish to run the $\mathrm{C}_{X}$ gate, we choose the phase where this curve is minimum so that $|11\rangle$ goes to $|10\rangle$.  The red curve shows the data when a control atom is not present and no blockade occurs and can therefore be thought of as $|01\rangle$.

\begin{figure}[!t] 
  \centering
  \includegraphics[width=.98\columnwidth]{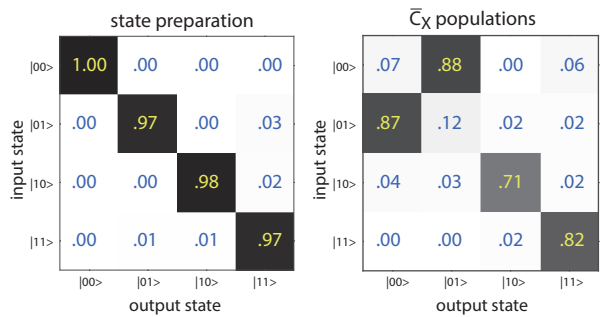}
  \caption{(color online) State preparation (left) and $\mathrm{\bar C}_{X}$ gate population matrix (right) on next nearest neighbor sites, 7.6~$\mu$m apart,  using parameters measured in Sec. \ref{sec:parmestimate}. The average statistical uncertainty of the data is $\pm 0.008$ for the state preparation and $\pm 0.02$ for the $\mathrm{\bar C}_{X}$ gate.}
  \label{fig:cnotgate}
\end{figure}

To measure the $\mathrm{\bar C}_{X}$ gate population matrix, each of the computational states are prepared, the $\mathrm{\bar C}_{X}$ pulse sequence is applied, and the results are  measured as shown in  Fig. \ref{fig:cnotgate}.  The overlap of the populations with an ideal gate is 0.82, without any corrections for atom loss. This improves on our previous result\cite{Zhang2010} of 0.74.
Note that the sum of the output populations in a given row is not equal to unity. For the first two rows we get 1.01, 1.03. We attribute this slight excess to fluctuations in the loading and atom retention rates. The second two rows have populations sums of 0.80, 0.84. For these input states $\ket{10}, \ket{11}$ the control qubit is Rydberg excited and must wait there while a Rydberg pulse is applied to the target qubit, before returning to the ground state. Excess loss of population in this case is the largest contributor to gate error. We discuss possible reasons for this loss in Sec. \ref{sec:discussion} below.

\subsection{Entanglement Results}\label{sec:bellExp}

\begin{figure}[!t]
  \centering
  \includegraphics[width=.99\columnwidth]{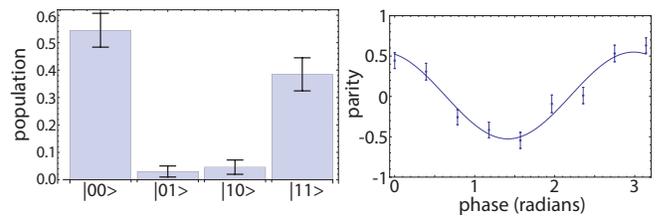}
  \caption{(color online) Bell state population measurement and parity oscillation measurement with no loss correction using next nearest neighbor sites.   The populations measured for the Bell state are $|00\rangle: 0.54\pm 0.06$, $|01\rangle: 0.03\pm 0.02$, $|10\rangle: 0.05\pm 0.03$, $|11\rangle: 0.38\pm 0.06$. The fit to the parity curve  gives  $|C_1|=0.27\pm0.02$ and $|C_2|=0.006$.  This results in an entanglement fidelity $F=0.73\pm0.05$ and  $F=0.79 \pm 0.05$ when corrected for atom loss.
}
  \label{fig:Bell82s}
\end{figure}

We proceed to implement the $U$ operator of Eq. (\ref{eq:bellstark}) 
to create an entangled two-qubit state. We select the gate phase of $\phi_{01}-\phi_{00}$ to implement the $\mathrm{\bar C}_{X}$ gate. 
Data are taken on next nearest neighbor sites, separated by 7.6~$\mu\rm m$. Parity oscillations are then performed to quantify the enaglement fidelity.    The results of Fig.~\ref{fig:Bell82s} give an entanglement fidelity, of $F=0.73\pm0.05$ without any loss correction.  Adding in a retention correction for ground state loss during the gate sequence  equal to 0.996 and 0.993 for the respective sites does not change this result.  Correcting for atom loss during the gate by re-normalizing each of the points on the parity curve  results in a loss corrected value for $C_1= 0.32 \pm 0.03$ and a post-selected entanglement fidelity of $F=0.79 \pm 0.05$. These results are the highest two-qubit entanglement fidelity, without post selection, reported to date using the Rydberg interaction. The previous best was 0.60\cite{Jau2015}.

\section{Discussion}\label{sec:discussion}

We have demonstrated improved entanglement fidelity using the Rydberg blockade interaction between two atomic qubits. Nevertheless the results obtained are still far  from  the $10^{-3}$, or lower, errors that are expected to be needed for scalable quantum computing\cite{Devitt2013}. Detailed calculations do predict the feasibility of much higher fidelity. It is therefore important to understand the cause of the observed infidelity and to implement improved protocols. 

One issue is the technical challenge of stabilizing all experimental parameters. Our current experimental procedures for qubit state measurements involve pushing out $f=4$ atoms, and then detecting the presence of an atom, in order to infer the qubit state. This method allows us to measure the qubit state with high
 fidelity\cite{Xia2015}, but implies that a new atom has to be loaded half the time on average. The need for atom reloading results in a relatively low data rate of 2 $\rm s^{-1}$. While state dependent measurements have been performed without atom 
loss\cite{Gibbons2011,*Fuhrmanek2011,Jau2015}, to date this has only been demonstrated on one or two trapped atoms. In the multi-site array  used here there is increased background noise from the trapping light and multiple atoms  which makes lossless measurements more difficult. Achieving lossless detection in the multi-site array is a challenge we are working to solve by increasing the optical detection efficiency.

\begin{figure*}[!t] 
  \centering
  \includegraphics[width=1.95 \columnwidth]{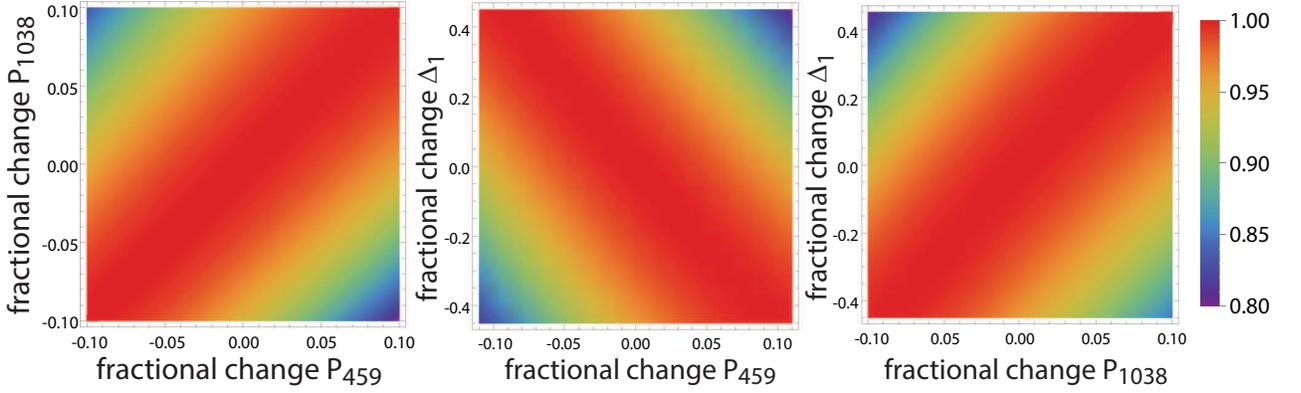}
  \caption{(color online) The entanglement fidelity  as a function of the fractional change of the optical powers $P_{459}, P_{1038}$ and the detuning $\Delta_1$ from  the optimal values.  The center point of each plot uses the parameters found experimentally in Sec. \ref{sec:parmestimate} and the color scale is the fidelity normalized by the value at the cener of each panel. }
  \label{fig:EntangleSens}
\vspace{-.3cm}
 \end{figure*}

The low data rate implies that entanglement experiments, including the requisite tuning of experimental parameters, as 
in Fig. \ref{fig.array}, require many hours to complete. As is shown in Fig. \ref{fig:EntangleSens} the fidelity drops steeply as experimental parameters are changed. In order to have a fidelity within 95\% of the optimal value the beam powers should not differ by more than $\pm 0.05$ and the detuning should not change by more than $\pm 0.2$. It is not difficult to maintain the detuning to the required precision, but holding beam intensities at the atoms to a few percent drift over many hours is challenging and needs to be improved on. 
Experimental intensity drifts are currently up to the 10\%  level over the course of a day  which contributes to the gate infidelity.

\begin{figure}[!t] 
  \centering
  \includegraphics[width=.99\columnwidth]{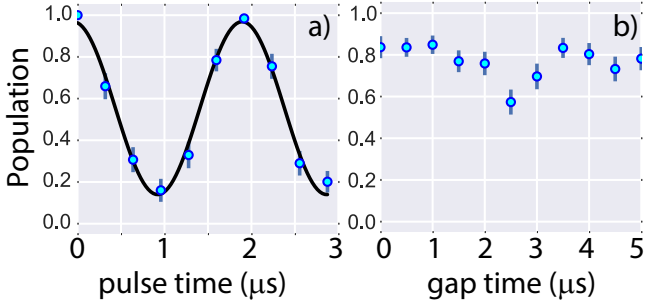}
  \caption{(color online) Ground state population after a $2\pi$ Rydberg pulse a) and after a $\pi-$ gap $-\pi$ Rydberg pulse sequence with variable gap time b).}
  \label{fig.pigappi}
\end{figure}

In addition to these technical issues the measured data point to a dominant experimental error. Looking at the last two rows in the $C_X$ population matrix in Fig. \ref{fig:cnotgate} we see that in cases where the control atom is Rydberg excited there is about 20\% loss of population.    This error, together with imperfect state preparation, and drifts of parameters at the few percent level easily explains the imperfect fidelity we obtain. Excess loss of Rydberg excited atoms has been the major contributor to gate infidelity not only in this study, but also in previous entanglement experiments \cite{Isenhower2010,Zhang2010,Wilk2010,Jau2015}. We note that for input state $\ket{01}$   where the  control atom is not excited, and the target atom experiences a $2\pi$ Rydberg pulse, the loss is very low, at most a few percent as shown in Fig. \ref{fig.pigappi}.
The difference between excitation of control or target atoms is that 
 the control atom has some gap time in between Rydberg $\pi$ rotations where the excitation beam is switched off, moved to the target site, then moved back to the control site, and switched back on for the final Rydberg $\pi$ rotation.  This gap time is equal to $\sim$~3~$\mu$s. This is much shorter than the lifetime of the Rydberg state and only slightly longer than the $2\pi$ Rydberg pulse applied to the target atom. We have verified in control experiments that a sequence of $\pi$ Rydberg pulse, very short gap time of $<50~\rm ns$, $\pi$ Rydberg pulse, also leads to excess Rydberg loss. 

The reasons for the additional loss when an atom is left in the Rydberg state for a short time are under investigation. Several possible explanations are worth considering. The large polarizability of Rydberg atoms could result in mechanical forces from background electric field gradients pushing the atoms away.  The experiments are performed in a pyrex vacuum cell with the atoms 1 cm away from the nearest walls. By way of spectroscopy on Rydberg states we have determined the background dc field to be $10-30 ~\rm mV/cm$. Although this measurement does not directly tell us about field gradients, assuming that the field varies by this amount over a length scale as short as 1 mm gives insufficient mechanical forces to explain the observed atom loss. Photoionization rates\cite{Saffman2005a} are also too small to explain the observed loss. 

Another possibility is that what we observe is not loss of an atom, but coupling between the laser excited $\ket{82s_{1/2},m_j=-1/2}$ state and some other Rydberg state during the gap time. Since other Rydberg states are not brought back to the ground state in the second 
$\pi$ pulse, such coupling could result in leaving an atom in the Rydberg state, and the atom then being lost when the trap light is turned back on after the gate sequence. Several mechanisms could result in coupling between different Rydberg levels. Although the $ns_{1/2}$ states have no significant tensor polarizability, matrix elements to other states are very large, scaling as $n^2 e a_0.$ Higher order terms in the hyperpolarizability\cite{Ilinova2009} could then lead to state mixing. Improved control of background electric fields should serve to limit this possibility. On the other hand if  a static background field was the reason for state mixing we might expect to see a similar loss in a single $2\pi$ Rydberg pulse, which we do not, as can be seen from Fig.  \ref{fig.pigappi}.

These considerations point to the likelihood that the observed loss is related to the turning on and off of the Rydberg pulses. The optical pulses are only approximations to square pulses, as they have  rise and fall times of about 20 ns. When there is 
a finite ground-Rydberg differential Stark shift there will be a time varying detuning at the rising and falling edges which may promote excitation of more than one Rydberg state. Figure \ref{fig.pigappi}b) shows a loss signal that does not increase monotonically with the gap time. Interestingly the minimal loss need not occur at the minimal gap time, and the population shows oscillatory  behavior.
Although we have  observed oscillations as a function of gap time, the details of the oscillations have not been repeatable.    The oscillations may be due to  time dependent interference of atomic states which are not eigenstates of the Rydberg excitation Hamiltonian. Numerical integration of the Schr\"odinger equation for such pulse sequences does  reveal  imperfect Rydberg excitation and de-excitation, and this issue is also the subject of ongoing work. 

Irrespective of the correct explanation for the excess Rydberg loss it is apparent that tuning the ground-Rydberg differential ac Stark shift to a value that is small compared to the excitation Rabi frequency has reduced the amount of loss, and improved the gate fidelity compared to previous work. We anticipate further improvement in future experiments with better control of background electric and magnetic fields as well as the use of optimized pulse shapes.

\acknowledgments

This research has been supported by the IARPA MQCO program through ARO contract No. W911NF-10-1-0347.

\appendix

\section{Rydberg excitation with intermediate state hyperfine structure}
\label{sec.appendixA}

The expressions given in the main text for the ground-Rydberg Rabi frequency and AC Stark shifts neglect the hyperfine structure of the intermediate level used for two-photon excitation. This is a good approximation when the detuning is very large compared to the width of the hyperfine structure.  
With our experimental parameters this approximation is only about 90\% accurate. We give here relevant formulae that account for the hyperfine structure.

We wish to couple atomic states $\ket{g}\rightarrow\ket{r}$ using two-photon excitation via intermediate level $\ket{p}.$ 
When the hyperfine structure of the intermediate level is negligible   the two-photon Rabi frequency is $\Omega_{\rm R}=\Omega_1\Omega_2/2\Delta$
with $\Omega_{1,2}$ the one photon Rabi frequencies and $\Delta=\omega_1-\omega_{pg}$ the detuning of the  field driving $\ket{g}\rightarrow \ket{p}.$ Here $\omega_{pg}=\omega_p - \omega_g=(U_p-U_g)/\hbar.$
This expression  is  accurate when $|\Delta|\gg \Delta_{\rm hf}$ where $\Delta_{\rm hf}$ is the width of the hyperfine structure of the $p$ level.

We proceed to calculate the Rabi frequency for two photon excitation $\ket{g}\stackrel{\mcE_{1}}{\rightarrow}\ket{p}
\stackrel{\mcE_{2}}{\rightarrow}\ket{r}$ where $\ket{g},\ket{r}$ are specific hyperfine states and $\ket{p}$ is shorthand for a manifold of hyperfine states. The two-photon matrix element between ground and Rydberg hyperfine states is 
\begin{widetext}
\bea
V_{f_g\rightarrow f_r}&=&
\mcE_1 \mcE_2 e^2
\bra{r,f_r,m_g+q_1+q_2}r_{q_2}\sum_{f_p}\ket{p,f_p,m_g+q_1}
\bra{p,f_p,m_g+q_1}r_{q_1}\ket{g,f_g,m_g}\nn\\
&=&V \sum_{f_p}
c_{Ij_gf_g}^{j_pf_p}c_{Ij_pf_p}^{j_rf_r}
C_{f_p,m_g+q_1,1,q_2}^{f_r,m_g+q_1+q_2}C_{f_g,m_g,1,q_1}^{f_p,m_g+q_1}
\eea
\end{widetext}
where 
$$
V=\mcE_1 \mcE_2 e^2\bra{n_rl_rsj_r}|r|\ket{n_pl_psj_p} \bra{n_pl_psj_p}|r|\ket{n_gl_gsj_g} , 
$$
 $C_{....}^{..}$ is a Clebsch-Gordan coefficient\cite{Varshalovich1988} and 
$$
c_{Ijf}^{j'f'}=(-1)^{1+I+f+j'}\sqrt{2f+1}S_{f'1j'}^{jIf},
$$
where $S_{def}^{abc}=\left\{\begin{matrix} a&b&c\\d&e&f\end{matrix}\right\}$ is a compact notation for the 6j symbol. Here $e$ is the electronic charge, $I$ is the nuclear spin, $s=1/2$ is the electronic spin of an alkali atom, $n_{g,p,r}$ is the principal quantum number of the atomic state, $f_{g,p,r}$ is the total angular momentum, $l_{g,p,r}$ is the orbital angular momentum, and $j_{g,p,r}$ is the electronic angular momentum. The photon fields 1 and 2 have amplitudes $\mcE_1, \mcE_2$, polarization state $q_1, q_2$ in a spherical basis and $m_g$ is the projection of the ground hyperfine state angular momentum. 

We define a one photon detuning from the center of mass of the  $p$ state
$\Delta_1= \omega_1-\omega_{pg}$ and hyperfine shifts within the $\ket{p}$ manifold $\Delta_{f_p}=\omega_{f_p}-\omega_{p}=2\pi\times \frac{A}{2}[f_p(f_p+1)-I(I+1)-j_p(j_p+1)]$ where $A$ is the magnetic dipole hyperfine constant. We only consider states with $j=1/2$ and can therefore neglect higher order terms of the hyperfine interaction.   

Adding the contributions from the $p$ levels the two-photon Rabi frequency between hyperfine states  can be written as 
\bea
\Omega_{f_g,m_g}^{f_r,m_g+q_1+q_2}=\Omega \tilde\Omega_{f_g,m_g}^{f_r,m_g+q_1+q_2}\nn
\eea
where
\begin{widetext}
\bea
\Omega&=&\frac{\mcE_1 \mcE_2 e^2\bra{n_rl_rsj_r}|r|\ket{n_pl_psj_p} \bra{n_pl_psj_p}|r|\ket{n_gl_gsj_g}}{2\hbar^2\Delta_1}
\nn\\
 \tilde\Omega_{f_g,m_g}^{f_r,m_g+q_1+q_2}&=&\sum_{f_p=|I-j_p|}^{I+j_p}
c_{Ij_gf_g}^{j_pf_p}c_{Ij_pf_p}^{j_rf_r}
C_{f_p,m_g+q_1,1,q_2}^{f_r,m_g+q_1+q_2}C_{f_g,m_g,1,q_1}^{f_p,m_g+q_1}
\frac{\Delta_1}{\Delta_1-\Delta_{f_p}}.\nn
\eea
\end{widetext}

The Rydberg hyperfine states can be expanded in an uncoupled basis as 
$$
\ket{j_rI;f_r,m_r}=\sum_{m_j,m_I}C_{j_rm_jIm_I}^{f_rm_r}\ket{j_rI;m_jm_I}
$$
or
$$
\ket{j_rI;m_jm_I}=\sum_{f_r,m_r}C_{j_rm_jIm_I}^{f_rm_r}\ket{j_rI;f_rm_r}.
$$
We are interested in the situation where we start in a single ground hyperfine state and therefore we  replace $m_r$  by the laser excited value $m_g+q_1+q_2$. 
Then since $\Omega_{f_g,m_g}^{f_r,m_r}=2\bra{f_rm_r}\opH\ket{f_gm_g}$, with $\opH$ the electric dipole Hamiltonian for the two-photon transition,  we can write the Rabi frequency coupling ground hyperfine and Rydberg fine structure states as 
\bea
\Omega_{f_g,m_g}^{j_r,m_{j}}&=&\sum_{f_r,m_r}C_{j_rm_jIm_I}^{f_rm_r}\Omega_{f_g,m_g}^{f_r,m_r}\nn\\
&=&\Omega\sum_{f_r}C_{j_rm_jIm_I}^{f_r,m_g+q_1+q_2}\tilde\Omega_{f_g,m_g}^{f_r,m_g+q_1+q_2}\nn\\
&\equiv& \Omega \bar\Omega_{f_g,m_g}^{j_r,m_j}  \label{eq.Rabihftofs}
\eea
with $m_I=m_{f_g}+q1+q2-m_{j_r}$. Here we have introduced an effective angular factor 
\be
\bar\Omega_{f_g,m_g}^{j_r,m_j}  =
\sum_{f_r}C_{j_r,m_j,I,m_g+q_1+q_2-m_{j}}^{f_r,m_g+q_1+q_2}\tilde\Omega_{f_g,m_g}^{f_r,m_g+q_1+q_2}.
\label{eq.OmegahfZ}
\ee
We use tildes to denote angular factors coupling hyperfine states to hyperfine states and overbars to denote factors coupling hyperfine states to fine structure Zeeman states. It should be emphasized that a description of the ground-Rydberg coupling in terms of a Rydberg fine structure state is only valid when the hyperfine interaction in the Rydberg state is negligible. Cesium $ns$ states have a relatively large hyperfine splitting so that in order to ensure coupling to a single Rydberg state we apply a bias magnetic field along $z$ to decouple the hyperfine interaction. This implies a modification to Eq. (\ref{eq.OmegahfZ}) which we will make explicit in Appendix \ref{sec.Rabi2fs}. 

We will also need the one-photon Rabi frequencies
\bea
\Xi_{f_g,m_{f_g},q_1}^{f_p}&=&\Xi_{gp} \tilde \Xi_{f_g,m_{f_g},q_1}^{f_p}\nn\\
&=& \Xi_{gp} c_{Ij_g f_g}^{j_p f_p}C_{f_g,m_{f_g},1,q_1}^{f_p,m_{f_g}+q_1}\nn\\
\Xi_{j_r,m_{j_r},-q_2}^{f_p}&=&\Xi_{rp} \bar \Xi_{j_r,m_{j_r},-q_2}^{f_p}\nn\\
&=& \Xi_{rp} \sum_{f_r}c_{I j_r f_r}^{j_p f_p}
C_{f_r,m_{f_r},1,-q_2}^{f_p,m_{f_r}-q_2} C_{j_r m_{j_r} I m_I}^{f_r, m_{f_r}}\nn
\eea
where $m_{f_r}=m_{f_g}+q_1+q_2$, $m_I= m_{f_r} - m_{j_r}$ and 
\bea
\Xi_{gp}&=&\frac{\mcE_1 e  \bra{n_pl_psj_p}|r|\ket{n_gl_gsj_g}}{\hbar},\nn\\
\Xi_{rp}&=&\frac{\mcE_2  e \bra{n_rl_rsj_r}|r|\ket{n_pl_psj_p}}{\hbar}\nn.
\eea
With these definitions the ground and Rydberg state Stark shifts under conditions of two-photon resonance are 
\bea
\Delta_{\rm ac, g}&=&\Xi_{gp}^2 \sum_{f_p} \frac{\left(\tilde\Xi_{f_g,m_{f_g},q_1}^{f_p}\right)^2}{4(\Delta_1-\Delta_{f_p})},\\
\Delta_{\rm ac, r}&=&\Xi_{rp}^2 
\sum_{f_p}\frac{\left(\bar\Xi_{j_r,m_{j_r},-q_2}^{f_p}\right)^2}{4(\Delta_1-\Delta_{f_p})}.
\label{eq.deltaR}
\eea
The differential AC Stark shift is $\Delta_{\rm ac} = \Delta_{\rm ac,r}-\Delta_{\rm ac,g}$. 

We have written the Rydberg fine structure state AC Stark shift as a sum over the contributions from hyperfine states of the $p$ level.
 When the Rydberg hyperfine coupling is not negligible we should instead use the hyperfine resolved AC Stark shifts which are 
\be 
\Delta_{\rm ac, f_r}=\Xi_{rp}^2 
\sum_{f_p}\frac{\left(\tilde\Xi_{f_r,m_{f_r},-q_2}^{f_p}\right)^2}{4(\Delta_1-\Delta_{f_p})}.\nn
\ee

Finally, it is also important to know the probability of photon scattering from the $p$ level 
during a ground to Rydberg $\pi$ pulse. The time for the pulse is $t_\pi =\pi/\Omega_{f_g,m_f}^{j_r,m_j}$ and the number of scattered photons is (see Sec. IV.B in \cite{Saffman2010})
\bea
N&=& \frac{\gamma_p t_\pi}{2}\left[\Xi_{gp}^2\sum_{f_p} \frac{\left(\tilde\Xi_{f_g,m_g,q_1}^{f_p}\right)^2}{2(\Delta_1-\Delta_{f_p})^2}+
\Xi_{rp}^2 \sum_{f_p}\frac{\left(\tilde\Xi_{j_r,m_j,-q_2}^{f_p}\right)^2}{2(\Delta_1-\Delta_{f_p})^2}\right]\nn.
\eea
Here $\gamma_p=1/\tau_p$ is the radiative decay rate from the $p$ level and the prefactor of $1/2$ acounts for half the integrated population being in the ground and Rydberg levels during the $\pi$ pulse. For coherent qubit control we choose parameters such that $N\ll 1$ and interpret $N$ as the probability $P_{\rm se}$ to scatter a photon. 

The above general expressions for Rabi frequency, AC Stark shift, and  spontaneous emission probability can be applied to any desired set of atomic levels. Expressions valid in  the limit of detuning large compared to the hyperfine structure width can be found by simply setting $\Delta_{f_p}\rightarrow 0.$

\section{Excitation of Cs $ns_{1/2}$ states via $7p_{1/2}$}
\label{sec.Rabi2fs}

Here we give explicit expressions for excitation of Cs atoms using $6s_{1/2}\rightarrow 7p_{1/2}\rightarrow ns_{1/2}$. The two fields are $\sigma_+, \sigma_-$ polarized so that the ground $\ket{6s_{1/2},f_g=4,m_f=0}$ state is coupled to the Rydberg hyperfine states $\ket{ns_{1/2},3,0}$, $\ket{ns_{1/2},4,0}$. The hyperfine-hyperfine angular factors defined in Appendix \ref{sec.appendixA} are 
\bea
\tilde\Xi_{4,0,1}^{3}&=&-\frac{1}{4},~~~\tilde\Xi_{3,0,1}^{3}=-\tilde\Xi_{4,0,1}^{3},\nn\\
\tilde\Xi_{4,0,1}^{4}&=&-\frac{\sqrt{5/3}}{4},~~~\tilde\Xi_{3,0,1}^{4}=-\tilde\Xi_{4,0,1}^{4},\nn\\
\tilde\Omega_{4,0}^{4,0}&=&\frac{1}{16}\frac{\Delta_1}{\Delta_1-\Delta_{3}}+\frac{5}{48}\frac{\Delta_1}{\Delta_1-\Delta_{4}},~~~
\tilde\Omega_{4,0}^{3,0}=-\tilde\Omega_{4,0}^{4,0}.\nn
\eea

For the experiments reported here we use Rydberg level $82s_{1/2}$ which has a hyperfine splitting between the $f_r=3,4$ components of about $\omega_{\rm hf,r}=2\pi\times 110.~\rm  kHz$. At zero magnetic field the $\ket{6s_{1/2},4,0}$ ground state couples to both hyperfine levels with equal and opposite Rabi frequencies. The non-zero hyperfine splitting implies that for any laser tuning there is some off-resonant excitation which leads to gate errors.

\begin{table}[!ht]
%\title{s states}
\setlength{\extrarowheight}{3pt}
\centering
\begin{tabular}{|l|c|c|}
\hline
parameter  & value& Ref.\\
\hline
$\alpha_{\rm g,459}^{\rm nr}$& $-11.6\times 10^{-24}~\left(\rm cm^3\right)$ & a) \\
$\alpha_{\rm g,1038}^{\rm nr}$& $189.\times 10^{-24}~\left(\rm cm^3\right)$ &a) \\
\hline
%$R_{6s(1/2)}^{7p(1/2)}$&$0.338\, a_0$\\
$\bra{7p_{1/2}}|r|\ket{6s_{1/2}}$&$-0.276\, a_0$&\cite{Vasilyev2002}\\
%$R_{7p(1/2)}^{ns(1/2)}$&$\frac{9.90}{n^{3/2}}\, a_0$\\
%$R_{7p(1/2)}^{nd(3/2)}$&$\frac{-19.5}{n^{3/2}}\, a_0$\\
$A_{7p_{1/2}}$&$ 94.35~\rm (MHz)$&\cite{Feiertag1972}\\
$\Delta_{3,{7p_{1/2}}}/(2\pi)$&$- 212.3 ~\rm (MHz)$&\\
$\Delta_{4,{7p_{1/2}}}/(2\pi)$&$  165.1~\rm (MHz)$&\\
$\tau_{7p_{1/2}}$& $0.155~(\mu\rm s)$&\cite{Ortiz1981} \\
\hline
$\alpha_{\rm 82s_{1/2},459}^{\rm nr}$& $-15.\times 10^{-24}~\left(\rm cm^3\right)$& Eq. (\ref{eq.alpharyd})\\
$\alpha_{\rm 82s_{1/2},1038}^{\rm nr}$& $-77.\times 10^{-24}~\left(\rm cm^3\right)$& Eq. (\ref{eq.alpharyd})\\
$\bra{ns_{1/2}}|r|\ket{7p_{1/2}}$&$\frac{-8.08}{n^{3/2}}\,a_0$&b) \\
%$\bra{nd_{3/2}}|r|\ket{7p_{1/2}}$&$\frac{-22.2}{n^{3/2}}\,a_0$\\
$A_{ns_{1/2}}$&$ \frac{13200.}{(n-4.05)^3}~\rm (MHz)$&c)\\
$\tau_{82s_{1/2}}$& $203.~(\mu\rm s) $&\cite*{Beterov2009,*Beterov2009b} \\
\hline
\end{tabular}
\caption{Physical parameters for Rydberg excitation of Cs via the $7p_{1/2}$ level. From top to bottom the table sections give ground state parameters, $7p_{1/2}$ parameters, and Rydberg level parameters. Reduced matrix elements are given in terms of the Bohr radius $a_0$.\\ a) The ground state nonresonant polarizabilities are calculated using a sum over states method,  excluding the $7p_{1/2}$ level in the case of $\alpha_{\rm g,459}^{\rm nr}$. \\
b) The $n$ dependence is a fit to values calculated using quantum defect wave functions as described in \cite{Walker2008}.\\
c) The $n$ dependence is an approximation based on values reported in \cite{Arimondo1977,Sassmannshausen2013}.  }
\label{tab.CsRyd1}
\end{table}

To correct for this we apply a small bias magnetic field $B_z$ along the $z$ axis of $B_z=0.15 ~\rm mT$. In the presence of the magnetic field the energies of the Rydberg $\ket{3,0}, \ket{4,0}$ states move apart and the coupled eigenstates can be written as 
\bea
\ket{82s_{1/2},m_j=1/2}&=& \frac{-\left(1-\sqrt{1+x^2} \right)\ket{3,0}+x \ket{4,0}}{\left[x^2+ \left(1-\sqrt{1+x^2} \right)^2\right]^{1/2}},\nn\\
\ket{82s_{1/2},m_j=-1/2}&=& \frac{\left(1+\sqrt{1+x^2} \right)\ket{3,0}-x \ket{4,0}}{\left[x^2+ \left(1+\sqrt{1+x^2} \right)^2\right]^{1/2}},\nn
\eea
with $x=\mu_B g_j B_z/\hbar\omega_{\rm r,hf}$ where $\mu_B$ is the Bohr magneton and $g_j\simeq2$. Here we have neglected the small correction due to the nuclear $g$ factor. 
At $B_z=0.15 ~\rm mT$ we find $x=38.2$ and the matrix element  from the ground state to  $\ket{82s_{1/2},m_j=-1/2}$ is within 0.0001 of the asymptotic value. At the same time the coupling to  $\ket{82s_{1/2},m_j=1/2}$ is suppressed to about 0.02 times the coupling to $\ket{4,0}$. The small matrix element together  with the approximately 4.2 MHz splitting between the $m_j=\pm 1/2$ states lets us tune to resonance with $m_j=-1/2$ and safely neglect the coupling to  $m_j=1/2.$ Measurements indicate a residual coupling to $m_j=1/2$ that is slightly larger than 0.02, which may be attributed to polarization errors of the Rydberg excitation beams.

\begin{figure*}[!t]
\begin{center}
\includegraphics[width=17.cm]{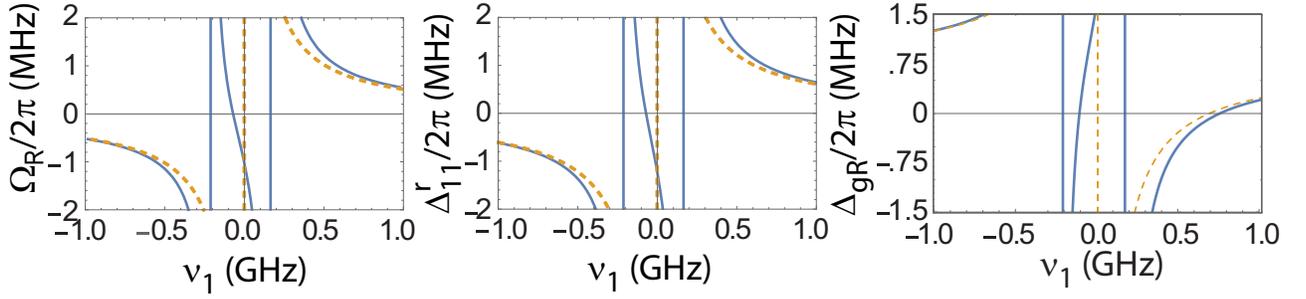}
\caption{(color online) Calculated ground-Rabi frequency $\Omega_{\rm R}$ (left), resonant Stark shift of $\ket{6s_{1/2},f=4}$ due to the 459 nm light $\Delta_{11}^{\rm r}$ (center), and total ground-Rydberg differential Stark shift $\Delta_{\rm gR}$ including resonant and non-resonant contributions  (right), for two-photon excitation of Cs $\ket{82s,m_j=-1/2}$ starting in $\ket{4,0}$ using parameters from Table  \ref{tab.expparameters} for the control site. Solid blue curves account for the $7p_{1/2}$ hyperfine structure, while the dashed yellow curves do not include hyperfine corrections.   The spontaneous emission plot only accounts for scattering from $7p_{1/2}$ as the Rydberg state decay is negligible for the parameters used. The gate experiments used $\nu_1=0.83 ~\rm GHz$.
 }
\label{fig.Cs82sRyd}
\end{center}
\end{figure*}

To summarize, with the magnetic field applied we couple to a Rydberg fine structure
state with matrix elements given by Eqs. (\ref{eq.Rabihftofs},\ref{eq.OmegahfZ},\ref{eq.deltaR}). 
Using the expressions in the previous section we find the ``reduced" one- and two-photon Rabi frequencies 
\bea
\frac{\Omega}{2\pi} &=& 87570. \times  \frac{(P_{1,\rm mW} P_{2\rm ,mW})^{1/2}}{n_r^{3/2} w_{1,\mu\rm m} w_{2,\mu\rm m} \nu_{1,\rm GHz}}~\rm (MHz),\nn\\
\frac{\Xi_{gp}}{2\pi}&=& 2446. \times  \frac{P_{1,\rm mW}^{1/2}}{ w_{1,\mu\rm m} }~\rm (MHz),\nn\\
\frac{\Xi_{rp}}{2\pi}&=&  71600.  \times \frac{P_{2,\rm mW}^{1/2}}{n_r^{3/2}  w_{2,\mu\rm m} }~\rm (MHz).\nn
\eea
where $\nu_{1,GHz}=\Delta_1/2\pi$ (in GHz), $P_{j, \rm mW}$ is a beam power in mW, and $w_{1(2),\mu\rm m}$ are beam waists ($1/e^2$ intensity radii) in $\mu\rm m$.  The two-photon angular factor coupling ground hyperfine  to  Rydberg fine structure state can be written compactly as   
\bea
\bar\Omega_{4,0}^{1/2,-1/2}&=& -\frac{1}{3\sqrt2 }\frac{1-\frac{5\Delta_{3,7p_{1/2}}}{8\Delta_{1}}-\frac{3\Delta_{4,7p_{1/2}}}{8\Delta_{1}}}
{\left(1-\frac{\Delta_{3,7p_{1/2}}}{\Delta_{1}}\right)\left(1-\frac{\Delta_{4,7p_{1/2}}}{\Delta_{1}}\right)}\nn
\eea
and $\bar\Omega_{3,0}^{1/2,-1/2}=-\bar\Omega_{4,0}^{1/2,-1/2}.$
Finally, the full expressions for the ground-Rydberg Rabi frequency, AC Stark shifts, and spontaneous emission probability from the $7p_{1/2}$ level are
\begin{widetext}
\bea
\frac{\Omega_{4,0}^{1/2,-1/2}}{2\pi}&=& -20600.  \frac{(P_{1\rm ,mW} P_{2,\rm mW})^{1/2}}{n_r^{3/2} w_{1,\mu\rm m} w_{2,\mu\rm m} \nu_{1,\rm GHz}} \frac{1-\frac{5\Delta_{3_p}}{8\Delta_{1}}-\frac{3\Delta_{4_p}}{8\Delta_{1}}}
{\left(1-\frac{\Delta_{3_p}}{\Delta_{1}}\right)\left(1-\frac{\Delta_{4_p}}{\Delta_{1}}\right)}  ~\rm (MHz)\nn,\\
\frac{\Delta_{\rm ac}}{2\pi}&=&\frac{\Delta_{\rm ac,r}-\Delta_{\rm ac,g}}{2\pi} \nn\\
&=&\left( 160200. \frac{P_{2,\rm mW}}{n_r^{3}  w_{2,\mu\rm m}^2 } -93.47 \frac{P_{1,\rm mW}}{ w_{1,\mu\rm m}^2 } \right) 
\frac{1}{\nu_{1,\rm GHz}}
\left(\frac{1}{1-\frac{\Delta_{3_p}}{\Delta_{1}}}
+\frac{5/3}{1-\frac{\Delta_{4_p}}{\Delta_{1}}}  \right) ~\rm (MHz)\nn,\\
P_{\rm se} &=& \left[ 0.43 \left(\frac{P_{2,\rm mW}}{n_r^3 P_{1,\rm mW}}\right)^{1/2}  \frac{ w_{1,\mu\rm m}}{  w_{2,\mu\rm m} }
+ 2.5\times 10^{-4} \left(\frac{n_r^3 P_{1,\rm mW}}{P_{2,\rm mW}}\right)^{1/2}   \frac{ w_{2,\mu\rm m}}{  w_{1,\mu\rm m} }   \right] \frac{1}{\nu_{1,\rm GHz}}\nn\\
&&\times 
\frac{1 - 2\left(\frac{5\Delta_{3_p}}{8\Delta_{1}}+ \frac{3\Delta_{4_p}}{8\Delta_{1}} \right)+\frac{5\Delta_{3_p}^2}{8\Delta_{1}^2}+ \frac{3\Delta_{4_p}^2}{8\Delta_{1}^2} }
{(1-\frac{\Delta_{3_p}}{\Delta_{1}})(1-\frac{\Delta_{4_p}}{\Delta_{1}})\left(1-\frac{5\Delta_{3_p}}{8\Delta_{1}}- \frac{3\Delta_{4_p}}{8\Delta_{1}} \right)}
\nn.
\eea
\end{widetext}

The ground-Rydberg Rabi frequency, resonant ground Stark shift, and total ground-Rydberg Stark shifts are shown versus detuning 
using experimental parameters in Fig. \ref{fig.Cs82sRyd}. Comparing the center and right panels demonstrates that the non-resonant Stark shifts must be accounted for when determining the parameters to be used to cancel the ground-Rydberg differential shift.

\bibliographystyle{apsrev4-1}

%\bibliography{d:/users/mark/pubs/biblio/saffman-refs,d:/users/mark/pubs/biblio/rydberg_bib_v15,d:/users/mark/pubs/biblio/qc_refs,d:/users/mark/pubs/biblio/atomic}	

\begin{thebibliography}{34}%
\makeatletter
\providecommand \@ifxundefined [1]{%
 \@ifx{#1\undefined}
}%
\providecommand \@ifnum [1]{%
 \ifnum #1\expandafter \@firstoftwo
 \else \expandafter \@secondoftwo
 \fi
}%
\providecommand \@ifx [1]{%
 \ifx #1\expandafter \@firstoftwo
 \else \expandafter \@secondoftwo
 \fi
}%
\providecommand \natexlab [1]{#1}%
\providecommand \enquote  [1]{``#1''}%
\providecommand \bibnamefont  [1]{#1}%
\providecommand \bibfnamefont [1]{#1}%
\providecommand \citenamefont [1]{#1}%
\providecommand \href@noop [0]{\@secondoftwo}%
\providecommand \href [0]{\begingroup \@sanitize@url \@href}%
\providecommand \@href[1]{\@@startlink{#1}\@@href}%
\providecommand \@@href[1]{\endgroup#1\@@endlink}%
\providecommand \@sanitize@url [0]{\catcode `\\12\catcode `\$12\catcode
  `\&12\catcode `\#12\catcode `\^12\catcode `\_12\catcode `\%12\relax}%
\providecommand \@@startlink[1]{}%
\providecommand \@@endlink[0]{}%
\providecommand \url  [0]{\begingroup\@sanitize@url \@url }%
\providecommand \@url [1]{\endgroup\@href {#1}{\urlprefix }}%
\providecommand \urlprefix  [0]{URL }%
\providecommand \Eprint [0]{\href }%
\providecommand \doibase [0]{http://dx.doi.org/}%
\providecommand \selectlanguage [0]{\@gobble}%
\providecommand \bibinfo  [0]{\@secondoftwo}%
\providecommand \bibfield  [0]{\@secondoftwo}%
\providecommand \translation [1]{[#1]}%
\providecommand \BibitemOpen [0]{}%
\providecommand \bibitemStop [0]{}%
\providecommand \bibitemNoStop [0]{.\EOS\space}%
\providecommand \EOS [0]{\spacefactor3000\relax}%
\providecommand \BibitemShut  [1]{\csname bibitem#1\endcsname}%
\let\auto@bib@innerbib\@empty
%</preamble>
\bibitem [{\citenamefont {Ladd}\ \emph {et~al.}(2010)\citenamefont {Ladd},
  \citenamefont {Jelezko}, \citenamefont {Laflamme}, \citenamefont {Nakamura},
  \citenamefont {Monroe},\ and\ \citenamefont {O'Brien}}]{Ladd2010}%
  \BibitemOpen
  \bibfield  {author} {\bibinfo {author} {\bibfnamefont {T.~D.}\ \bibnamefont
  {Ladd}}, \bibinfo {author} {\bibfnamefont {F.}~\bibnamefont {Jelezko}},
  \bibinfo {author} {\bibfnamefont {R.}~\bibnamefont {Laflamme}}, \bibinfo
  {author} {\bibfnamefont {Y.}~\bibnamefont {Nakamura}}, \bibinfo {author}
  {\bibfnamefont {C.}~\bibnamefont {Monroe}}, \ and\ \bibinfo {author}
  {\bibfnamefont {J.~L.}\ \bibnamefont {O'Brien}},\ }\href@noop {} {\bibfield
  {journal} {\bibinfo  {journal} {Nature}\ }\textbf {\bibinfo {volume} {464}},\
  \bibinfo {pages} {45} (\bibinfo {year} {2010})}\BibitemShut {NoStop}%
\bibitem [{\citenamefont {Xia}\ \emph {et~al.}(2015)\citenamefont {Xia},
  \citenamefont {Lichtman}, \citenamefont {Maller}, \citenamefont {Carr},
  \citenamefont {Piotrowicz}, \citenamefont {Isenhower},\ and\ \citenamefont
  {Saffman}}]{Xia2015}%
  \BibitemOpen
  \bibfield  {author} {\bibinfo {author} {\bibfnamefont {T.}~\bibnamefont
  {Xia}}, \bibinfo {author} {\bibfnamefont {M.}~\bibnamefont {Lichtman}},
  \bibinfo {author} {\bibfnamefont {K.}~\bibnamefont {Maller}}, \bibinfo
  {author} {\bibfnamefont {A.~W.}\ \bibnamefont {Carr}}, \bibinfo {author}
  {\bibfnamefont {M.~J.}\ \bibnamefont {Piotrowicz}}, \bibinfo {author}
  {\bibfnamefont {L.}~\bibnamefont {Isenhower}}, \ and\ \bibinfo {author}
  {\bibfnamefont {M.}~\bibnamefont {Saffman}},\ }\href@noop {} {\bibfield
  {journal} {\bibinfo  {journal} {Phys. Rev. Lett.}\ }\textbf {\bibinfo
  {volume} {114}},\ \bibinfo {pages} {100503} (\bibinfo {year}
  {2015})}\BibitemShut {NoStop}%
\bibitem [{\citenamefont {Jaksch}\ \emph {et~al.}(2000)\citenamefont {Jaksch},
  \citenamefont {Cirac}, \citenamefont {Zoller}, \citenamefont {Rolston},
  \citenamefont {C\^ot\'e},\ and\ \citenamefont {Lukin}}]{Jaksch2000}%
  \BibitemOpen
  \bibfield  {author} {\bibinfo {author} {\bibfnamefont {D.}~\bibnamefont
  {Jaksch}}, \bibinfo {author} {\bibfnamefont {J.~I.}\ \bibnamefont {Cirac}},
  \bibinfo {author} {\bibfnamefont {P.}~\bibnamefont {Zoller}}, \bibinfo
  {author} {\bibfnamefont {S.~L.}\ \bibnamefont {Rolston}}, \bibinfo {author}
  {\bibfnamefont {R.}~\bibnamefont {C\^ot\'e}}, \ and\ \bibinfo {author}
  {\bibfnamefont {M.~D.}\ \bibnamefont {Lukin}},\ }\href@noop {} {\bibfield
  {journal} {\bibinfo  {journal} {Phys. Rev. Lett.}\ }\textbf {\bibinfo
  {volume} {85}},\ \bibinfo {pages} {2208} (\bibinfo {year}
  {2000})}\BibitemShut {NoStop}%
\bibitem [{\citenamefont {Saffman}\ and\ \citenamefont
  {Walker}(2005)}]{Saffman2005a}%
  \BibitemOpen
  \bibfield  {author} {\bibinfo {author} {\bibfnamefont {M.}~\bibnamefont
  {Saffman}}\ and\ \bibinfo {author} {\bibfnamefont {T.~G.}\ \bibnamefont
  {Walker}},\ }\href@noop {} {\bibfield  {journal} {\bibinfo  {journal} {Phys.
  Rev. A}\ }\textbf {\bibinfo {volume} {72}},\ \bibinfo {pages} {022347}
  (\bibinfo {year} {2005})}\BibitemShut {NoStop}%
\bibitem [{\citenamefont {Piotrowicz}\ \emph {et~al.}(2013)\citenamefont
  {Piotrowicz}, \citenamefont {Lichtman}, \citenamefont {Maller}, \citenamefont
  {Li}, \citenamefont {Zhang}, \citenamefont {Isenhower},\ and\ \citenamefont
  {Saffman}}]{Piotrowicz2013}%
  \BibitemOpen
  \bibfield  {author} {\bibinfo {author} {\bibfnamefont {M.~J.}\ \bibnamefont
  {Piotrowicz}}, \bibinfo {author} {\bibfnamefont {M.}~\bibnamefont
  {Lichtman}}, \bibinfo {author} {\bibfnamefont {K.}~\bibnamefont {Maller}},
  \bibinfo {author} {\bibfnamefont {G.}~\bibnamefont {Li}}, \bibinfo {author}
  {\bibfnamefont {S.}~\bibnamefont {Zhang}}, \bibinfo {author} {\bibfnamefont
  {L.}~\bibnamefont {Isenhower}}, \ and\ \bibinfo {author} {\bibfnamefont
  {M.}~\bibnamefont {Saffman}},\ }\href@noop {} {\bibfield  {journal} {\bibinfo
   {journal} {Phys. Rev. A}\ }\textbf {\bibinfo {volume} {88}},\ \bibinfo
  {pages} {013420} (\bibinfo {year} {2013})}\BibitemShut {NoStop}%
\bibitem [{\citenamefont {Devitt}\ \emph {et~al.}(2013)\citenamefont {Devitt},
  \citenamefont {Munro},\ and\ \citenamefont {Nemoto}}]{Devitt2013}%
  \BibitemOpen
  \bibfield  {author} {\bibinfo {author} {\bibfnamefont {S.~J.}\ \bibnamefont
  {Devitt}}, \bibinfo {author} {\bibfnamefont {W.~J.}\ \bibnamefont {Munro}}, \
  and\ \bibinfo {author} {\bibfnamefont {K.}~\bibnamefont {Nemoto}},\
  }\href@noop {} {\bibfield  {journal} {\bibinfo  {journal} {Rep. Prog. Phys.}\
  }\textbf {\bibinfo {volume} {76}},\ \bibinfo {pages} {076001} (\bibinfo
  {year} {2013})}\BibitemShut {NoStop}%
\bibitem [{\citenamefont {Merrill}\ \emph {et~al.}(2014)\citenamefont
  {Merrill}, \citenamefont {Doret}, \citenamefont {Vittorini}, \citenamefont
  {Addison},\ and\ \citenamefont {Brown}}]{Merrill2014}%
  \BibitemOpen
  \bibfield  {author} {\bibinfo {author} {\bibfnamefont {J.~T.}\ \bibnamefont
  {Merrill}}, \bibinfo {author} {\bibfnamefont {S.~C.}\ \bibnamefont {Doret}},
  \bibinfo {author} {\bibfnamefont {G.}~\bibnamefont {Vittorini}}, \bibinfo
  {author} {\bibfnamefont {J.~P.}\ \bibnamefont {Addison}}, \ and\ \bibinfo
  {author} {\bibfnamefont {K.~R.}\ \bibnamefont {Brown}},\ }\href@noop {}
  {\bibfield  {journal} {\bibinfo  {journal} {Phys. Rev. A}\ }\textbf {\bibinfo
  {volume} {90}},\ \bibinfo {pages} {040301(R)} (\bibinfo {year}
  {2014})}\BibitemShut {NoStop}%
\bibitem [{\citenamefont {Mount}\ \emph {et~al.}(2015)\citenamefont {Mount},
  \citenamefont {Kabytayev}, \citenamefont {Crain}, \citenamefont {Harper},
  \citenamefont {Baek}, \citenamefont {Vrijsen}, \citenamefont {Flammia},
  \citenamefont {Brown}, \citenamefont {Maunz},\ and\ \citenamefont
  {Kim}}]{Mount2015}%
  \BibitemOpen
  \bibfield  {author} {\bibinfo {author} {\bibfnamefont {E.}~\bibnamefont
  {Mount}}, \bibinfo {author} {\bibfnamefont {C.}~\bibnamefont {Kabytayev}},
  \bibinfo {author} {\bibfnamefont {S.}~\bibnamefont {Crain}}, \bibinfo
  {author} {\bibfnamefont {R.}~\bibnamefont {Harper}}, \bibinfo {author}
  {\bibfnamefont {S.-Y.}\ \bibnamefont {Baek}}, \bibinfo {author}
  {\bibfnamefont {G.}~\bibnamefont {Vrijsen}}, \bibinfo {author} {\bibfnamefont
  {S.}~\bibnamefont {Flammia}}, \bibinfo {author} {\bibfnamefont {K.~R.}\
  \bibnamefont {Brown}}, \bibinfo {author} {\bibfnamefont {P.}~\bibnamefont
  {Maunz}}, \ and\ \bibinfo {author} {\bibfnamefont {J.}~\bibnamefont {Kim}},\
  }\href@noop {} {\bibfield  {journal} {\bibinfo  {journal} {arXiv:1504.01440}\
  } (\bibinfo {year} {2015})}\BibitemShut {NoStop}%
\bibitem [{\citenamefont {Isenhower}\ \emph {et~al.}(2010)\citenamefont
  {Isenhower}, \citenamefont {Urban}, \citenamefont {Zhang}, \citenamefont
  {Gill}, \citenamefont {Henage}, \citenamefont {Johnson}, \citenamefont
  {Walker},\ and\ \citenamefont {Saffman}}]{Isenhower2010}%
  \BibitemOpen
  \bibfield  {author} {\bibinfo {author} {\bibfnamefont {L.}~\bibnamefont
  {Isenhower}}, \bibinfo {author} {\bibfnamefont {E.}~\bibnamefont {Urban}},
  \bibinfo {author} {\bibfnamefont {X.~L.}\ \bibnamefont {Zhang}}, \bibinfo
  {author} {\bibfnamefont {A.~T.}\ \bibnamefont {Gill}}, \bibinfo {author}
  {\bibfnamefont {T.}~\bibnamefont {Henage}}, \bibinfo {author} {\bibfnamefont
  {T.~A.}\ \bibnamefont {Johnson}}, \bibinfo {author} {\bibfnamefont {T.~G.}\
  \bibnamefont {Walker}}, \ and\ \bibinfo {author} {\bibfnamefont
  {M.}~\bibnamefont {Saffman}},\ }\href@noop {} {\bibfield  {journal} {\bibinfo
   {journal} {Phys. Rev. Lett.}\ }\textbf {\bibinfo {volume} {104}},\ \bibinfo
  {pages} {010503} (\bibinfo {year} {2010})}\BibitemShut {NoStop}%
\bibitem [{\citenamefont {Zhang}\ \emph {et~al.}(2010)\citenamefont {Zhang},
  \citenamefont {Isenhower}, \citenamefont {Gill}, \citenamefont {Walker},\
  and\ \citenamefont {Saffman}}]{Zhang2010}%
  \BibitemOpen
  \bibfield  {author} {\bibinfo {author} {\bibfnamefont {X.~L.}\ \bibnamefont
  {Zhang}}, \bibinfo {author} {\bibfnamefont {L.}~\bibnamefont {Isenhower}},
  \bibinfo {author} {\bibfnamefont {A.~T.}\ \bibnamefont {Gill}}, \bibinfo
  {author} {\bibfnamefont {T.~G.}\ \bibnamefont {Walker}}, \ and\ \bibinfo
  {author} {\bibfnamefont {M.}~\bibnamefont {Saffman}},\ }\href@noop {}
  {\bibfield  {journal} {\bibinfo  {journal} {Phys. Rev. A}\ }\textbf {\bibinfo
  {volume} {82}},\ \bibinfo {pages} {030306(R)} (\bibinfo {year}
  {2010})}\BibitemShut {NoStop}%
\bibitem [{\citenamefont {Wilk}\ \emph {et~al.}(2010)\citenamefont {Wilk},
  \citenamefont {Ga\"etan}, \citenamefont {Evellin}, \citenamefont {Wolters},
  \citenamefont {Miroshnychenko}, \citenamefont {Grangier},\ and\ \citenamefont
  {Browaeys}}]{Wilk2010}%
  \BibitemOpen
  \bibfield  {author} {\bibinfo {author} {\bibfnamefont {T.}~\bibnamefont
  {Wilk}}, \bibinfo {author} {\bibfnamefont {A.}~\bibnamefont {Ga\"etan}},
  \bibinfo {author} {\bibfnamefont {C.}~\bibnamefont {Evellin}}, \bibinfo
  {author} {\bibfnamefont {J.}~\bibnamefont {Wolters}}, \bibinfo {author}
  {\bibfnamefont {Y.}~\bibnamefont {Miroshnychenko}}, \bibinfo {author}
  {\bibfnamefont {P.}~\bibnamefont {Grangier}}, \ and\ \bibinfo {author}
  {\bibfnamefont {A.}~\bibnamefont {Browaeys}},\ }\href@noop {} {\bibfield
  {journal} {\bibinfo  {journal} {Phys. Rev. Lett.}\ }\textbf {\bibinfo
  {volume} {104}},\ \bibinfo {pages} {010502} (\bibinfo {year}
  {2010})}\BibitemShut {NoStop}%
\bibitem [{\citenamefont {Jau}\ \emph {et~al.}(2015)\citenamefont {Jau},
  \citenamefont {Hankin}, \citenamefont {Keating}, \citenamefont {Deutsch},\
  and\ \citenamefont {Biedermann}}]{Jau2015}%
  \BibitemOpen
  \bibfield  {author} {\bibinfo {author} {\bibfnamefont {Y.-Y.}\ \bibnamefont
  {Jau}}, \bibinfo {author} {\bibfnamefont {A.~M.}\ \bibnamefont {Hankin}},
  \bibinfo {author} {\bibfnamefont {T.}~\bibnamefont {Keating}}, \bibinfo
  {author} {\bibfnamefont {I.~H.}\ \bibnamefont {Deutsch}}, \ and\ \bibinfo
  {author} {\bibfnamefont {G.~W.}\ \bibnamefont {Biedermann}},\ }\href@noop {}
  {\bibfield  {journal} {\bibinfo  {journal} {arXiv:1501.03862}\ } (\bibinfo
  {year} {2015})}\BibitemShut {NoStop}%
\bibitem [{\citenamefont {Zhang}\ \emph {et~al.}(2012)\citenamefont {Zhang},
  \citenamefont {Gill}, \citenamefont {Isenhower}, \citenamefont {Walker},\
  and\ \citenamefont {Saffman}}]{XZhang2012}%
  \BibitemOpen
  \bibfield  {author} {\bibinfo {author} {\bibfnamefont {X.~L.}\ \bibnamefont
  {Zhang}}, \bibinfo {author} {\bibfnamefont {A.~T.}\ \bibnamefont {Gill}},
  \bibinfo {author} {\bibfnamefont {L.}~\bibnamefont {Isenhower}}, \bibinfo
  {author} {\bibfnamefont {T.~G.}\ \bibnamefont {Walker}}, \ and\ \bibinfo
  {author} {\bibfnamefont {M.}~\bibnamefont {Saffman}},\ }\href@noop {}
  {\bibfield  {journal} {\bibinfo  {journal} {Phys. Rev. A}\ }\textbf {\bibinfo
  {volume} {85}},\ \bibinfo {pages} {042310} (\bibinfo {year}
  {2012})}\BibitemShut {NoStop}%
\bibitem [{\citenamefont {Xia}\ \emph {et~al.}(2013)\citenamefont {Xia},
  \citenamefont {Zhang},\ and\ \citenamefont {Saffman}}]{Xia2013}%
  \BibitemOpen
  \bibfield  {author} {\bibinfo {author} {\bibfnamefont {T.}~\bibnamefont
  {Xia}}, \bibinfo {author} {\bibfnamefont {X.~L.}\ \bibnamefont {Zhang}}, \
  and\ \bibinfo {author} {\bibfnamefont {M.}~\bibnamefont {Saffman}},\
  }\href@noop {} {\bibfield  {journal} {\bibinfo  {journal} {Phys. Rev. A}\
  }\textbf {\bibinfo {volume} {88}},\ \bibinfo {pages} {062337} (\bibinfo
  {year} {2013})}\BibitemShut {NoStop}%
\bibitem [{\citenamefont {Tong}\ \emph {et~al.}(2004)\citenamefont {Tong},
  \citenamefont {Farooqi}, \citenamefont {Stanojevic}, \citenamefont
  {Krishnan}, \citenamefont {Zhang}, \citenamefont {C\^ot\'e}, \citenamefont
  {Eyler},\ and\ \citenamefont {Gould}}]{Tong2004}%
  \BibitemOpen
  \bibfield  {author} {\bibinfo {author} {\bibfnamefont {D.}~\bibnamefont
  {Tong}}, \bibinfo {author} {\bibfnamefont {S.~M.}\ \bibnamefont {Farooqi}},
  \bibinfo {author} {\bibfnamefont {J.}~\bibnamefont {Stanojevic}}, \bibinfo
  {author} {\bibfnamefont {S.}~\bibnamefont {Krishnan}}, \bibinfo {author}
  {\bibfnamefont {Y.~P.}\ \bibnamefont {Zhang}}, \bibinfo {author}
  {\bibfnamefont {R.}~\bibnamefont {C\^ot\'e}}, \bibinfo {author}
  {\bibfnamefont {E.~E.}\ \bibnamefont {Eyler}}, \ and\ \bibinfo {author}
  {\bibfnamefont {P.~L.}\ \bibnamefont {Gould}},\ }\href@noop {} {\bibfield
  {journal} {\bibinfo  {journal} {Phys. Rev. Lett.}\ }\textbf {\bibinfo
  {volume} {93}},\ \bibinfo {pages} {063001} (\bibinfo {year}
  {2004})}\BibitemShut {NoStop}%
\bibitem [{\citenamefont {Manthey}\ \emph {et~al.}(2014)\citenamefont
  {Manthey}, \citenamefont {Weber}, \citenamefont {Niederpr\"um}, \citenamefont
  {Langer}, \citenamefont {Guarrera}, \citenamefont {Barontini},\ and\
  \citenamefont {Ott}}]{Manthey2014}%
  \BibitemOpen
  \bibfield  {author} {\bibinfo {author} {\bibfnamefont {T.}~\bibnamefont
  {Manthey}}, \bibinfo {author} {\bibfnamefont {T.~M.}\ \bibnamefont {Weber}},
  \bibinfo {author} {\bibfnamefont {T.}~\bibnamefont {Niederpr\"um}}, \bibinfo
  {author} {\bibfnamefont {P.}~\bibnamefont {Langer}}, \bibinfo {author}
  {\bibfnamefont {V.}~\bibnamefont {Guarrera}}, \bibinfo {author}
  {\bibfnamefont {G.}~\bibnamefont {Barontini}}, \ and\ \bibinfo {author}
  {\bibfnamefont {H.}~\bibnamefont {Ott}},\ }\href@noop {} {\bibfield
  {journal} {\bibinfo  {journal} {New J. Phys.}\ }\textbf {\bibinfo {volume}
  {16}},\ \bibinfo {pages} {083034} (\bibinfo {year} {2014})}\BibitemShut
  {NoStop}%
\bibitem [{\citenamefont {Hankin}\ \emph {et~al.}(2014)\citenamefont {Hankin},
  \citenamefont {Jau}, \citenamefont {Parazzoli}, \citenamefont {Chou},
  \citenamefont {Armstrong}, \citenamefont {Landahl},\ and\ \citenamefont
  {Biedermann}}]{Hankin2014}%
  \BibitemOpen
  \bibfield  {author} {\bibinfo {author} {\bibfnamefont {A.~M.}\ \bibnamefont
  {Hankin}}, \bibinfo {author} {\bibfnamefont {Y.-Y.}\ \bibnamefont {Jau}},
  \bibinfo {author} {\bibfnamefont {L.~P.}\ \bibnamefont {Parazzoli}}, \bibinfo
  {author} {\bibfnamefont {C.~W.}\ \bibnamefont {Chou}}, \bibinfo {author}
  {\bibfnamefont {D.~J.}\ \bibnamefont {Armstrong}}, \bibinfo {author}
  {\bibfnamefont {A.~J.}\ \bibnamefont {Landahl}}, \ and\ \bibinfo {author}
  {\bibfnamefont {G.~W.}\ \bibnamefont {Biedermann}},\ }\href@noop {}
  {\bibfield  {journal} {\bibinfo  {journal} {Phys. Rev. A}\ }\textbf {\bibinfo
  {volume} {89}},\ \bibinfo {pages} {033416} (\bibinfo {year}
  {2014})}\BibitemShut {NoStop}%
\bibitem [{\citenamefont {Sa\ss{}mannshausen}\ \emph
  {et~al.}(2013)\citenamefont {Sa\ss{}mannshausen}, \citenamefont {Merkt},\
  and\ \citenamefont {Deiglmayr}}]{Sassmannshausen2013}%
  \BibitemOpen
  \bibfield  {author} {\bibinfo {author} {\bibfnamefont {H.}~\bibnamefont
  {Sa\ss{}mannshausen}}, \bibinfo {author} {\bibfnamefont {F.}~\bibnamefont
  {Merkt}}, \ and\ \bibinfo {author} {\bibfnamefont {J.}~\bibnamefont
  {Deiglmayr}},\ }\href@noop {} {\bibfield  {journal} {\bibinfo  {journal}
  {Phys. Rev. A}\ }\textbf {\bibinfo {volume} {87}},\ \bibinfo {pages} {032519}
  (\bibinfo {year} {2013})}\BibitemShut {NoStop}%
\bibitem [{\citenamefont {Zhang}\ \emph {et~al.}(2011)\citenamefont {Zhang},
  \citenamefont {Robicheaux},\ and\ \citenamefont {Saffman}}]{SZhang2011}%
  \BibitemOpen
  \bibfield  {author} {\bibinfo {author} {\bibfnamefont {S.}~\bibnamefont
  {Zhang}}, \bibinfo {author} {\bibfnamefont {F.}~\bibnamefont {Robicheaux}}, \
  and\ \bibinfo {author} {\bibfnamefont {M.}~\bibnamefont {Saffman}},\
  }\href@noop {} {\bibfield  {journal} {\bibinfo  {journal} {Phys. Rev. A}\
  }\textbf {\bibinfo {volume} {84}},\ \bibinfo {pages} {043408} (\bibinfo
  {year} {2011})}\BibitemShut {NoStop}%
\bibitem [{\citenamefont {Turchette}\ \emph {et~al.}(1998)\citenamefont
  {Turchette}, \citenamefont {Wood}, \citenamefont {King}, \citenamefont
  {Myatt}, \citenamefont {Leibfried}, \citenamefont {Itano}, \citenamefont
  {Monroe},\ and\ \citenamefont {Wineland}}]{Turchette1998}%
  \BibitemOpen
  \bibfield  {author} {\bibinfo {author} {\bibfnamefont {Q.~A.}\ \bibnamefont
  {Turchette}}, \bibinfo {author} {\bibfnamefont {C.~S.}\ \bibnamefont {Wood}},
  \bibinfo {author} {\bibfnamefont {B.~E.}\ \bibnamefont {King}}, \bibinfo
  {author} {\bibfnamefont {C.~J.}\ \bibnamefont {Myatt}}, \bibinfo {author}
  {\bibfnamefont {D.}~\bibnamefont {Leibfried}}, \bibinfo {author}
  {\bibfnamefont {W.~M.}\ \bibnamefont {Itano}}, \bibinfo {author}
  {\bibfnamefont {C.}~\bibnamefont {Monroe}}, \ and\ \bibinfo {author}
  {\bibfnamefont {D.~J.}\ \bibnamefont {Wineland}},\ }\href@noop {} {\bibfield
  {journal} {\bibinfo  {journal} {Phys. Rev. Lett.}\ }\textbf {\bibinfo
  {volume} {81}},\ \bibinfo {pages} {3631} (\bibinfo {year}
  {1998})}\BibitemShut {NoStop}%
\bibitem [{\citenamefont {Sackett}\ \emph {et~al.}(2000)\citenamefont
  {Sackett}, \citenamefont {Kielpinski}, \citenamefont {King}, \citenamefont
  {Langer}, \citenamefont {Meyer}, \citenamefont {Myatt}, \citenamefont {Rowe},
  \citenamefont {Turchette}, \citenamefont {Itano}, \citenamefont {Wineland},\
  and\ \citenamefont {Monroe}}]{Sackett2000}%
  \BibitemOpen
  \bibfield  {author} {\bibinfo {author} {\bibfnamefont {C.~A.}\ \bibnamefont
  {Sackett}}, \bibinfo {author} {\bibfnamefont {D.}~\bibnamefont {Kielpinski}},
  \bibinfo {author} {\bibfnamefont {B.~E.}\ \bibnamefont {King}}, \bibinfo
  {author} {\bibfnamefont {C.}~\bibnamefont {Langer}}, \bibinfo {author}
  {\bibfnamefont {V.}~\bibnamefont {Meyer}}, \bibinfo {author} {\bibfnamefont
  {C.~J.}\ \bibnamefont {Myatt}}, \bibinfo {author} {\bibfnamefont
  {M.}~\bibnamefont {Rowe}}, \bibinfo {author} {\bibfnamefont {Q.~A.}\
  \bibnamefont {Turchette}}, \bibinfo {author} {\bibfnamefont {W.~M.}\
  \bibnamefont {Itano}}, \bibinfo {author} {\bibfnamefont {D.~J.}\ \bibnamefont
  {Wineland}}, \ and\ \bibinfo {author} {\bibfnamefont {C.}~\bibnamefont
  {Monroe}},\ }\href@noop {} {\bibfield  {journal} {\bibinfo  {journal} {Nature
  (London)}\ }\textbf {\bibinfo {volume} {404}},\ \bibinfo {pages} {256}
  (\bibinfo {year} {2000})}\BibitemShut {NoStop}%
\bibitem [{\citenamefont {Maller}\ \emph {et~al.}(2015)\citenamefont {Maller},
  \citenamefont {Lichtman},\ and\ \citenamefont {Saffman}}]{Maller2015b}%
  \BibitemOpen
  \bibfield  {author} {\bibinfo {author} {\bibfnamefont {K.}~\bibnamefont
  {Maller}}, \bibinfo {author} {\bibfnamefont {M.}~\bibnamefont {Lichtman}}, \
  and\ \bibinfo {author} {\bibfnamefont {M.}~\bibnamefont {Saffman}},\
  }\href@noop {} {\bibfield  {journal} {\bibinfo  {journal} {in preparation}\ }
  (\bibinfo {year} {2015})}\BibitemShut {NoStop}%
\bibitem [{\citenamefont {Gibbons}\ \emph {et~al.}(2011)\citenamefont
  {Gibbons}, \citenamefont {Hamley}, \citenamefont {Shih},\ and\ \citenamefont
  {Chapman}}]{Gibbons2011}%
  \BibitemOpen
  \bibfield  {author} {\bibinfo {author} {\bibfnamefont {M.~J.}\ \bibnamefont
  {Gibbons}}, \bibinfo {author} {\bibfnamefont {C.~D.}\ \bibnamefont {Hamley}},
  \bibinfo {author} {\bibfnamefont {C.-Y.}\ \bibnamefont {Shih}}, \ and\
  \bibinfo {author} {\bibfnamefont {M.~S.}\ \bibnamefont {Chapman}},\
  }\href@noop {} {\bibfield  {journal} {\bibinfo  {journal} {Phys. Rev. Lett.}\
  }\textbf {\bibinfo {volume} {106}},\ \bibinfo {pages} {133002} (\bibinfo
  {year} {2011})}\BibitemShut {NoStop}%
\bibitem [{\citenamefont {Fuhrmanek}\ \emph {et~al.}(2011)\citenamefont
  {Fuhrmanek}, \citenamefont {Bourgain}, \citenamefont {Sortais},\ and\
  \citenamefont {Browaeys}}]{Fuhrmanek2011}%
  \BibitemOpen
  \bibfield  {author} {\bibinfo {author} {\bibfnamefont {A.}~\bibnamefont
  {Fuhrmanek}}, \bibinfo {author} {\bibfnamefont {R.}~\bibnamefont {Bourgain}},
  \bibinfo {author} {\bibfnamefont {Y.~R.~P.}\ \bibnamefont {Sortais}}, \ and\
  \bibinfo {author} {\bibfnamefont {A.}~\bibnamefont {Browaeys}},\ }\href@noop
  {} {\bibfield  {journal} {\bibinfo  {journal} {Phys. Rev. Lett.}\ }\textbf
  {\bibinfo {volume} {106}},\ \bibinfo {pages} {133003} (\bibinfo {year}
  {2011})}\BibitemShut {NoStop}%
\bibitem [{\citenamefont {Il'inova}\ \emph {et~al.}(2009)\citenamefont
  {Il'inova}, \citenamefont {Kamenski},\ and\ \citenamefont
  {Ovsiannikov}}]{Ilinova2009}%
  \BibitemOpen
  \bibfield  {author} {\bibinfo {author} {\bibfnamefont {E.~Y.}\ \bibnamefont
  {Il'inova}}, \bibinfo {author} {\bibfnamefont {A.~A.}\ \bibnamefont
  {Kamenski}}, \ and\ \bibinfo {author} {\bibfnamefont {V.~D.}\ \bibnamefont
  {Ovsiannikov}},\ }\href@noop {} {\bibfield  {journal} {\bibinfo  {journal}
  {J. Phys. B: At. Mol. Opt. Phys.}\ }\textbf {\bibinfo {volume} {42}},\
  \bibinfo {pages} {145004} (\bibinfo {year} {2009})}\BibitemShut {NoStop}%
\bibitem [{\citenamefont {Varshalovich}\ \emph {et~al.}(1988)\citenamefont
  {Varshalovich}, \citenamefont {Moskalev},\ and\ \citenamefont
  {Khersonskii}}]{Varshalovich1988}%
  \BibitemOpen
  \bibfield  {author} {\bibinfo {author} {\bibfnamefont {D.~A.}\ \bibnamefont
  {Varshalovich}}, \bibinfo {author} {\bibfnamefont {A.~N.}\ \bibnamefont
  {Moskalev}}, \ and\ \bibinfo {author} {\bibfnamefont {V.~K.}\ \bibnamefont
  {Khersonskii}},\ }\href@noop {} {\emph {\bibinfo {title} {Quantum theory of
  angular momentum}}}\ (\bibinfo  {publisher} {World Scientific},\ \bibinfo
  {year} {1988})\BibitemShut {NoStop}%
\bibitem [{\citenamefont {Saffman}\ \emph {et~al.}(2010)\citenamefont
  {Saffman}, \citenamefont {Walker},\ and\ \citenamefont
  {M\o{}lmer}}]{Saffman2010}%
  \BibitemOpen
  \bibfield  {author} {\bibinfo {author} {\bibfnamefont {M.}~\bibnamefont
  {Saffman}}, \bibinfo {author} {\bibfnamefont {T.~G.}\ \bibnamefont {Walker}},
  \ and\ \bibinfo {author} {\bibfnamefont {K.}~\bibnamefont {M\o{}lmer}},\
  }\href@noop {} {\bibfield  {journal} {\bibinfo  {journal} {Rev. Mod. Phys.}\
  }\textbf {\bibinfo {volume} {82}},\ \bibinfo {pages} {2313} (\bibinfo {year}
  {2010})}\BibitemShut {NoStop}%
\bibitem [{\citenamefont {Vasilyev}\ \emph {et~al.}(2002)\citenamefont
  {Vasilyev}, \citenamefont {Savukov}, \citenamefont {Safronova},\ and\
  \citenamefont {Berry}}]{Vasilyev2002}%
  \BibitemOpen
  \bibfield  {author} {\bibinfo {author} {\bibfnamefont {A.~A.}\ \bibnamefont
  {Vasilyev}}, \bibinfo {author} {\bibfnamefont {I.~M.}\ \bibnamefont
  {Savukov}}, \bibinfo {author} {\bibfnamefont {M.~S.}\ \bibnamefont
  {Safronova}}, \ and\ \bibinfo {author} {\bibfnamefont {H.~G.}\ \bibnamefont
  {Berry}},\ }\href@noop {} {\bibfield  {journal} {\bibinfo  {journal} {Phys.
  Rev. A}\ }\textbf {\bibinfo {volume} {66}},\ \bibinfo {pages} {020101}
  (\bibinfo {year} {2002})}\BibitemShut {NoStop}%
\bibitem [{\citenamefont {Feiertag}\ \emph {et~al.}(1972)\citenamefont
  {Feiertag}, \citenamefont {Sahm},\ and\ \citenamefont
  {zu~Putlitz}}]{Feiertag1972}%
  \BibitemOpen
  \bibfield  {author} {\bibinfo {author} {\bibfnamefont {D.}~\bibnamefont
  {Feiertag}}, \bibinfo {author} {\bibfnamefont {A.}~\bibnamefont {Sahm}}, \
  and\ \bibinfo {author} {\bibfnamefont {G.}~\bibnamefont {zu~Putlitz}},\
  }\href@noop {} {\bibfield  {journal} {\bibinfo  {journal} {Z. Physik}\
  }\textbf {\bibinfo {volume} {255}},\ \bibinfo {pages} {93} (\bibinfo {year}
  {1972})}\BibitemShut {NoStop}%
\bibitem [{\citenamefont {Ortiz}\ and\ \citenamefont
  {Campos}(1981)}]{Ortiz1981}%
  \BibitemOpen
  \bibfield  {author} {\bibinfo {author} {\bibfnamefont {M.}~\bibnamefont
  {Ortiz}}\ and\ \bibinfo {author} {\bibfnamefont {J.}~\bibnamefont {Campos}},\
  }\href@noop {} {\bibfield  {journal} {\bibinfo  {journal} {J. Quant.
  Spectros. Rad. Transf.}\ }\textbf {\bibinfo {volume} {26}},\ \bibinfo {pages}
  {107} (\bibinfo {year} {1981})}\BibitemShut {NoStop}%
\bibitem [{\citenamefont {Beterov}\ \emph
  {et~al.}(2009{\natexlab{a}})\citenamefont {Beterov}, \citenamefont
  {Ryabtsev}, \citenamefont {Tretyakov},\ and\ \citenamefont
  {Entin}}]{Beterov2009}%
  \BibitemOpen
  \bibfield  {author} {\bibinfo {author} {\bibfnamefont {I.~I.}\ \bibnamefont
  {Beterov}}, \bibinfo {author} {\bibfnamefont {I.~I.}\ \bibnamefont
  {Ryabtsev}}, \bibinfo {author} {\bibfnamefont {D.~B.}\ \bibnamefont
  {Tretyakov}}, \ and\ \bibinfo {author} {\bibfnamefont {V.~M.}\ \bibnamefont
  {Entin}},\ }\href@noop {} {\bibfield  {journal} {\bibinfo  {journal} {Phys.
  Rev. A}\ }\textbf {\bibinfo {volume} {79}},\ \bibinfo {pages} {052504}
  (\bibinfo {year} {2009}{\natexlab{a}})}\BibitemShut {NoStop}%
\bibitem [{\citenamefont {Beterov}\ \emph
  {et~al.}(2009{\natexlab{b}})\citenamefont {Beterov}, \citenamefont
  {Ryabtsev}, \citenamefont {Tretyakov},\ and\ \citenamefont
  {Entin}}]{Beterov2009b}%
  \BibitemOpen
  \bibfield  {author} {\bibinfo {author} {\bibfnamefont {I.~I.}\ \bibnamefont
  {Beterov}}, \bibinfo {author} {\bibfnamefont {I.~I.}\ \bibnamefont
  {Ryabtsev}}, \bibinfo {author} {\bibfnamefont {D.~B.}\ \bibnamefont
  {Tretyakov}}, \ and\ \bibinfo {author} {\bibfnamefont {V.~M.}\ \bibnamefont
  {Entin}},\ }\href@noop {} {\bibfield  {journal} {\bibinfo  {journal} {Phys.
  Rev. A}\ }\textbf {\bibinfo {volume} {80}},\ \bibinfo {pages} {059902}
  (\bibinfo {year} {2009}{\natexlab{b}})}\BibitemShut {NoStop}%
\bibitem [{\citenamefont {Walker}\ and\ \citenamefont
  {Saffman}(2008)}]{Walker2008}%
  \BibitemOpen
  \bibfield  {author} {\bibinfo {author} {\bibfnamefont {T.~G.}\ \bibnamefont
  {Walker}}\ and\ \bibinfo {author} {\bibfnamefont {M.}~\bibnamefont
  {Saffman}},\ }\href@noop {} {\bibfield  {journal} {\bibinfo  {journal} {Phys.
  Rev. A}\ }\textbf {\bibinfo {volume} {77}},\ \bibinfo {pages} {032723}
  (\bibinfo {year} {2008})}\BibitemShut {NoStop}%
\bibitem [{\citenamefont {Arimondo}\ \emph {et~al.}(1977)\citenamefont
  {Arimondo}, \citenamefont {Inguscio},\ and\ \citenamefont
  {Violino}}]{Arimondo1977}%
  \BibitemOpen
  \bibfield  {author} {\bibinfo {author} {\bibfnamefont {E.}~\bibnamefont
  {Arimondo}}, \bibinfo {author} {\bibfnamefont {M.}~\bibnamefont {Inguscio}},
  \ and\ \bibinfo {author} {\bibfnamefont {P.}~\bibnamefont {Violino}},\
  }\href@noop {} {\bibfield  {journal} {\bibinfo  {journal} {Rev. Mod. Phys.}\
  }\textbf {\bibinfo {volume} {49}},\ \bibinfo {pages} {31} (\bibinfo {year}
  {1977})}\BibitemShut {NoStop}%
\end{thebibliography}

%merlin.mbs apsrev4-1.bst 2010-07-25 4.21a (PWD, AO, DPC) hacked
%Control: key (0)
%Control: author (72) initials jnrlst
%Control: editor formatted (1) identically to author
%Control: production of article title (-1) disabled
%Control: page (0) single
%Control: year (1) truncated
%Control: production of eprint (0) enabled
%

\end{document}